\documentclass[reprint,preprintnumbers,superscriptaddress,amsmath,amssymb,aps]{revtex4-1}

\usepackage{graphicx}
\usepackage{dcolumn}
\usepackage{bm}
\usepackage{varwidth}
\usepackage[export]{adjustbox}
\usepackage{threeparttable}
\usepackage{multirow}
\usepackage[usenames, dvipsnames]{color}
\usepackage[mathlines]{lineno}

\graphicspath{{figs/}}

\begin{document}

\title{Suppressing parametric instabilities in LIGO using low-noise acoustic mode dampers}

\author{S.~Biscans}
\affiliation {LIGO, Massachusetts Institute of Technology, Cambridge, MA 02139, USA }
\affiliation {LIGO, California Institute of Technology, Pasadena, CA 91125, USA  }
\author{S.~Gras\thanks{Corresponding author}}
\email{sgras@ligo.mit.edu}
\affiliation {LIGO, Massachusetts Institute of Technology, Cambridge, MA 02139, USA }
\author{C.D.~Blair}
\affiliation {LIGO Livingston Observatory, Livingston, LA 70754, USA }
\author{J.~Driggers}
\affiliation {LIGO Hanford Observatory, Richland, WA 99354, USA }      
\author{M.~Evans}
\affiliation {LIGO, Massachusetts Institute of Technology, Cambridge, MA 02139, USA }
\author{P.~Fritschel}
\affiliation {LIGO, Massachusetts Institute of Technology, Cambridge, MA 02139, USA }
\author{T.~Hardwick}
\affiliation {Louisiana State University, Baton Rouge, LA 70803, USA }
\author{G.~Mansell}
\affiliation {LIGO Hanford Observatory, Richland, WA 99354, USA }      

\begin{abstract}

Interferometric gravitational-wave detectors like LIGO need to be able to measure changes in their arm lengths of order $10^{-18}~$m or smaller. This requires very high laser power in order to raise the signal above shot noise. One significant limitation to increased laser power is an opto-mechanical interaction between the laser field and the detector's test masses that can form an unstable feedback loop. Such parametric instabilities have long been studied as a limiting effect at high power, and were first observed to occur in LIGO in 2014. Since then, passive and active means have been used to avoid these instabilities, though at power levels well below the final design value. Here we report on the successful implementation of tuned, passive dampers to tame parametric instabilities in LIGO. These dampers are applied directly to all interferometer test masses to reduce the quality factors of their internal vibrational modes, while adding a negligible amount of noise to the gravitational-wave output. In accordance with our model, the measured mode quality factors have been reduced by at least a factor of ten with no visible increase in the interferometer's thermal noise level. We project that these dampers should remove most of the parametric instabilities in LIGO when operating at full power, while limiting the concomitant increase in thermal noise to approximately 1\%.

\end{abstract}

\maketitle


\section{Introduction} \label{introduction}
\subsection{Overview}

\begin{figure*}[t]
\centering
\includegraphics[width=\textwidth]{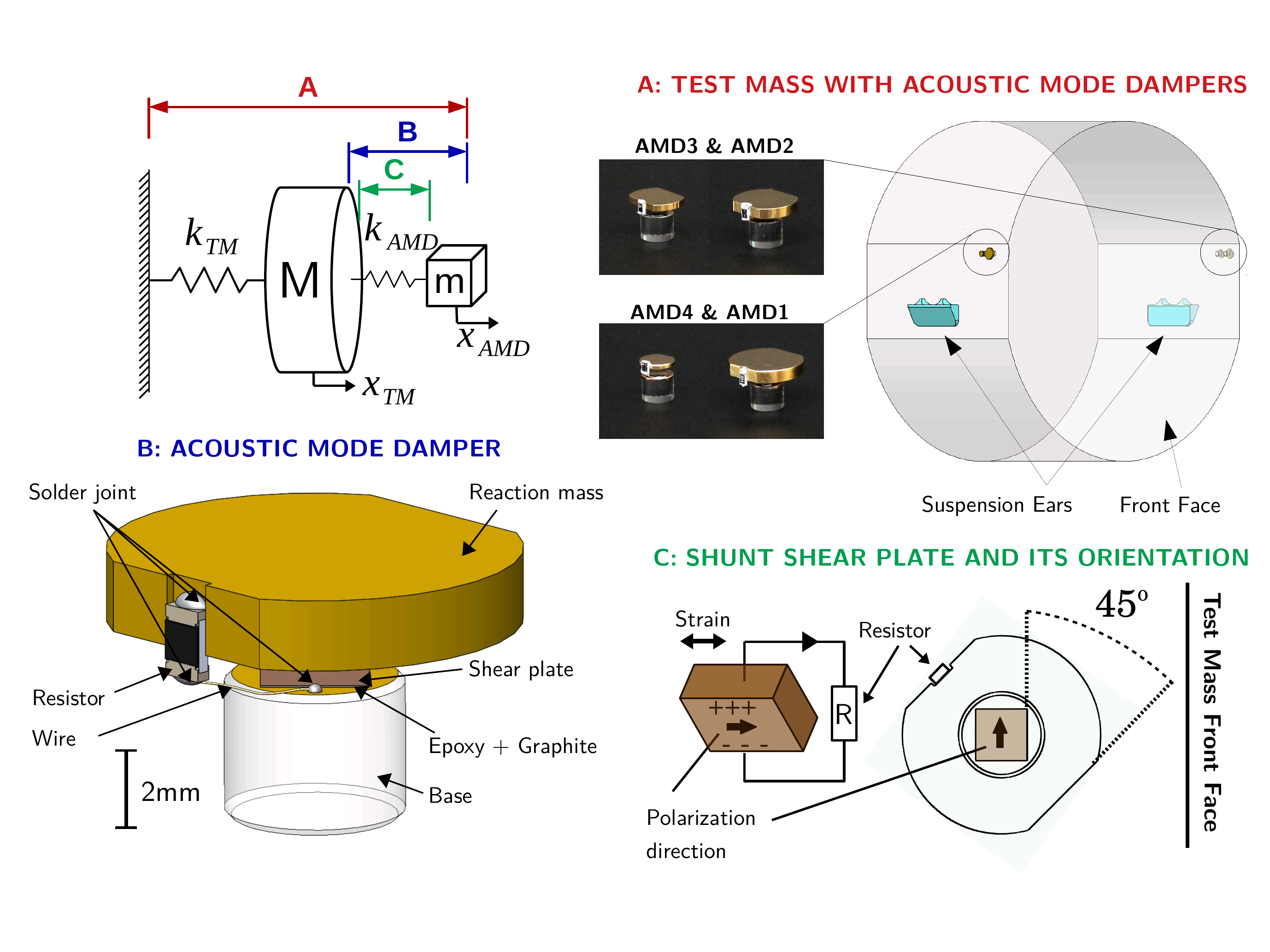}
\caption{Overview of the low-noise Acoustic Mode Damper. The AMD can be described as a small damper of mass $m$ attached to a larger vibrating mass $M$, as illustrated in the top left. To cover a broader frequency bandwidth, each test mass is equipped with four different AMDs distributed on the optic's flats, as shown in Fig.~A. Each AMD is made of a base, a shunted shear plate and a reaction mass (Fig.~B). The shunted shear plate is used as a lossy tunable spring with a complex stiffness $k_{AMD}$. Its polarization direction is oriented perpendicular to the cavity axis to limit thermal noise injection (Fig.~C). Finally, the top face of the base and the entire reaction mass are gold coated for electrical conductivity, assuring current flow between the PZT plate and the resistor. The bonds with the PZT plate are made of epoxy mixed with graphite nano-particles for conductivity. A detailed description of AMD components can be found in Table~\ref{table:param}.}
\label{fig:amd}
\end{figure*}

Interferometric gravitational-wave detectors use a modified Michelson interferometer that measures gravitational-wave strain as a difference in length of its orthogonal arms, which are made several kilometers long to increase their strain-to-length conversion. Other enhancements to the basic Michelson interferometer are made to increase the conversion of path length change to optical signal. These include the use of resonant optical cavities in the long arms to multiply the light phase change, an input power-recycling mirror that creates additional resonant buildup of the laser light in the interferometer, and an output signal-recycling mirror that broadens the bandwidth of the arm cavities. The quantum-noise limited sensitivity of the interferometer is determined by the stored laser power, and, up to a limit, is improved by increasing the laser power. For the 11 gravitational-wave detections made in their first two observation runs~\cite{ligo2018gwtc}, the Advanced LIGO interferometers operated with 100-120$\,$kW of power stored in each arm cavity. 
Since the full design sensitivity of Advanced LIGO calls for 750$\,$kW of arm power~\cite{aasi2015advanced},
higher power will be required to reach the instruments' full potential.

There are however significant technical challenges to achieving and maintaining stable operation 
as the laser power is increased. One of these involves opto-mechanical interactions between the stored laser field and the arm cavities’ test masses that can form an unstable feedback
loop~\cite{evans2010general}. Given the high optical power level in each cavity  
and the very high mechanical quality factors ($Q$-factors) of the test mass vibrational modes ($\gtrsim 10^7$), the process can result in a parametric instability (PI), in which the cavity optical energy is pumped into a test mass mechanical mode, which grows exponentially until the interferometer becomes inoperable. 

Since Braginsky \textit{et al.}~\cite{braginsky2001parametric} identified the phenomenon, PI have been extensively studied as a limitation for advanced interferometric gravitational-wave detectors \cite{braginsky2002analysis,kells2002considerations,ju2006multiple,gras2010parametric,evans2010general,vyatchanin2012parametric}.  A PI was first observed in early operation of the Advanced LIGO interferometers, where a 15.5$\,$kHz test mass mode interacted with a third-order transverse optical mode of an arm cavity, exhibiting unstable growth when the arm power exceeded 25$\,$kW~\cite{evans2015observation}. With 100$\,$kW of arm cavity power, several modes were potentially unstable in each detector.

In the first two observing runs, these unstable modes were suppressed with one of two techniques. The first PI 
was stabilized by shifting the eigenfrequency of the third-order optical mode to reduce the 
optical gain at the mechanical mode frequency~\cite{zhao2005parametric}. This was 
done by thermally decreasing the radius-of-curvature of one of the cavity test masses, using 
a non-contacting radiative heater that surrounds the barrel of each test mass. Unstable modes
have also been suppressed actively, using feedback forces applied to the test masses to effectively reduce their internal mode $Q$-factors~\cite{blair2017first}.

At full power, approximately 10 modes in each test mass would be unstable if not otherwise mitigated~\cite{evans2015observation},
and neither of these techniques are expected to be robust at that level. In the thermal tuning technique, thermally shifting the optical higher-order mode spacing can decrease the optical gain for some modes, but it will increase the optical gain for other modes that will eventually become unstable. The active damping approach becomes complicated in the face of dozens of modes to damp, some of which are very close in frequency. Each requires a suitable sensing signal and careful signal processing to avoid interactions between modes; it can quickly become a game of whack-a-mole.

A third approach is to reduce the test mass $Q$-factors passively, with the application of some type of damping mechanism. 
The challenge of this approach is to provide adequate damping in the (15-80)$\,$kHz band, while minimally impacting the test mass thermal 
noise around 100$\,$Hz, in order to preserve the detector’s strain sensitivity. This means the dampers must add negligible mechanical loss at frequencies 
well below their resonances.
Gras \textit{et al.}~\cite{gras2010parametric} investigated the use of metal rings and coatings applied to the circumference of the test mass, but they
could achieve appreciable damping of the $Q$-factors only by adding enough damping material that the test mass thermal noise was increased significantly.
A more frequency-selective damper was required, which led to the idea of tuned dampers designed to resonantly damp modes in critical frequency band of
(15-80)$\,$kHz~\cite{gras2015resonant}. The prototype acoustic mode damper (AMD) reported in~\cite{gras2015resonant} showed promising performance in terms of 
mode damping, but was estimated to more than double the thermal noise at 100$\,$Hz if applied to the test mass and thus was also not a practical design. 

In this article we present a new design of a much lower-noise AMD, suitable for application in advanced gravitational-wave detectors. The basic design of the AMD remains the
same as that presented in ~\cite{gras2015resonant}, but each element of the damper has been modified and optimized to reduce its noise impact. These AMDs have been 
applied to all four test masses of both LIGO interferometers. 
The resulting measured $Q$-factors are roughly an order of magnitude smaller than without the dampers, consistent with our model predictions. 
The AMDs are enabling instability-free operation in the (15-80)$\,$kHz band during Advanced LIGO's third observation run (O3) at up to 30\% of full power. We project
that all modes should remain stable at or close to full power operation in that frequency band. The estimated degradation of the Advanced LIGO design strain noise due to 
the AMDs is at most 1.0\%, and we present a measurement that is consistent with this projection.


\subsection{Parametric Instabilities}

The process that leads to PI can be viewed as a closed-loop feedback mechanism~\cite{evans2010general} involving interactions between 3 modes: the fundamental optical mode of the arm cavity (Hermite-Gaussian ${\rm TEM_{00}}$ mode); a higher-order transverse optical mode of the arm cavity; and an internal vibrational mode of a test mass. Feedback occurs when the cavity fundamental mode reflects from a test mass surface that is vibrating at a mechanical eigenmode (due merely to thermal excitation, e.g.), scattering a very small fraction of the fundamental mode into higher-order optical modes in the cavity. Via radiation pressure, the beat note of the fundamental and higher-order optical modes exert a spatially varying force on the cavity test masses, which oscillates at the mechanical mode frequency. This force can further drive the amplitude of the mechanical mode, closing the loop. Depending on the frequency relationship between the mechanical and optical modes, the feedback may be positive or negative. 

The dynamics of this process are commonly described in terms of the parametric gain $R$, with
$R > 1$ being the threshold for instability. The parametric gain for a mechanical mode $m$ with eigenfrequency $f_m$ can be expressed as:

\begin{equation}
R_{m} = \frac{2 P_{arm}}{\pi \lambda c}\frac{Q_{m} }{M f_{m}^2 } \sum_{n=0}^{\infty} Re[G_{n}]B^{2}_{m,n} ,
\label{eq:rm}
\end{equation}
where $P_{arm}$ is the laser power stored in the cavity, $M$ is the mass of the test mass, $c$ is
the speed of light, $\lambda$ is the laser
wavelength, and $Q_m$ is the $Q$-factor of the mechanical mode. The factor $B_{m,n}$ is the geometrical overlap between the mechanical mode $m$ and an optical mode $n$, and $Re[G_{n}]$ is the real part of the optical gain for mode $n$. The summation is over all higher-order optical modes which can contribute to the $R_{m}$ value, though typically 
only one is relevant~\cite{evans2015observation}. 

The amplitude of the mode $m$ is governed by an exponential, $e^{t/\tau_{pi}}$, with the time
constant
\begin{equation}
\tau_{pi} = \frac{\tau_{m}}{(R_{m}-1)} \, ,
\label{eq:tau}
\end{equation}
where $\tau_{m}$ is the natural decay time of the mechanical mode $m$ in the absence of the
opto-mechanical interaction, and is related to the $Q$-factor as $Q_{m} = \pi \tau_{m} f_{m}$.

For $R_{m} < 1$, the time constant is negative and the mode amplitude decays exponentially, at a rate that may be longer or shorter than the natural decay time.  
For values of $R_{m} > 1$, $\tau_{pi}$ is positive, indicating exponential growth of the mechanical mode. The parametric
gain scales linearly with $P_{arm}$ and $Q_m$, and the strategy behind the AMD is to lower the $Q$-factors so that $R$ stays below unity for all modes.
 

\section{Low-Noise Acoustic Mode Damper Concept}

Tuned mass damping is a well-established technique for controlling mechanical vibrations \cite{losurdo1999inverted, ballardin2001measurement,matichard2015advanced, bergmann2017passive}, and piezoelectric tuned mass dampers are being developed as energy harvesting devices \cite{sodano2004estimation}. Designing tuned dampers for the test masses of a gravitational-wave detector presents 
unique challenges, as they must not only provide broadband $Q$-reduction in the  PI band (15-80) kHz, but they must also preserve the inherently low mechanical loss 
of the test mass in the gravitational-wave band to maintain a low level of thermal noise in that band. 

In this section we first describe the AMD and its interaction with the test mass with a simple one-dimensional model in order to illustrate the concept
and its feasibility. Then we show how the specific design is optimized using a complete finite-element model of the entire system.

\subsection{One-dimensional model of the AMD}

\begin{figure}[h!]
	\centering
	\includegraphics[width=0.5\textwidth]{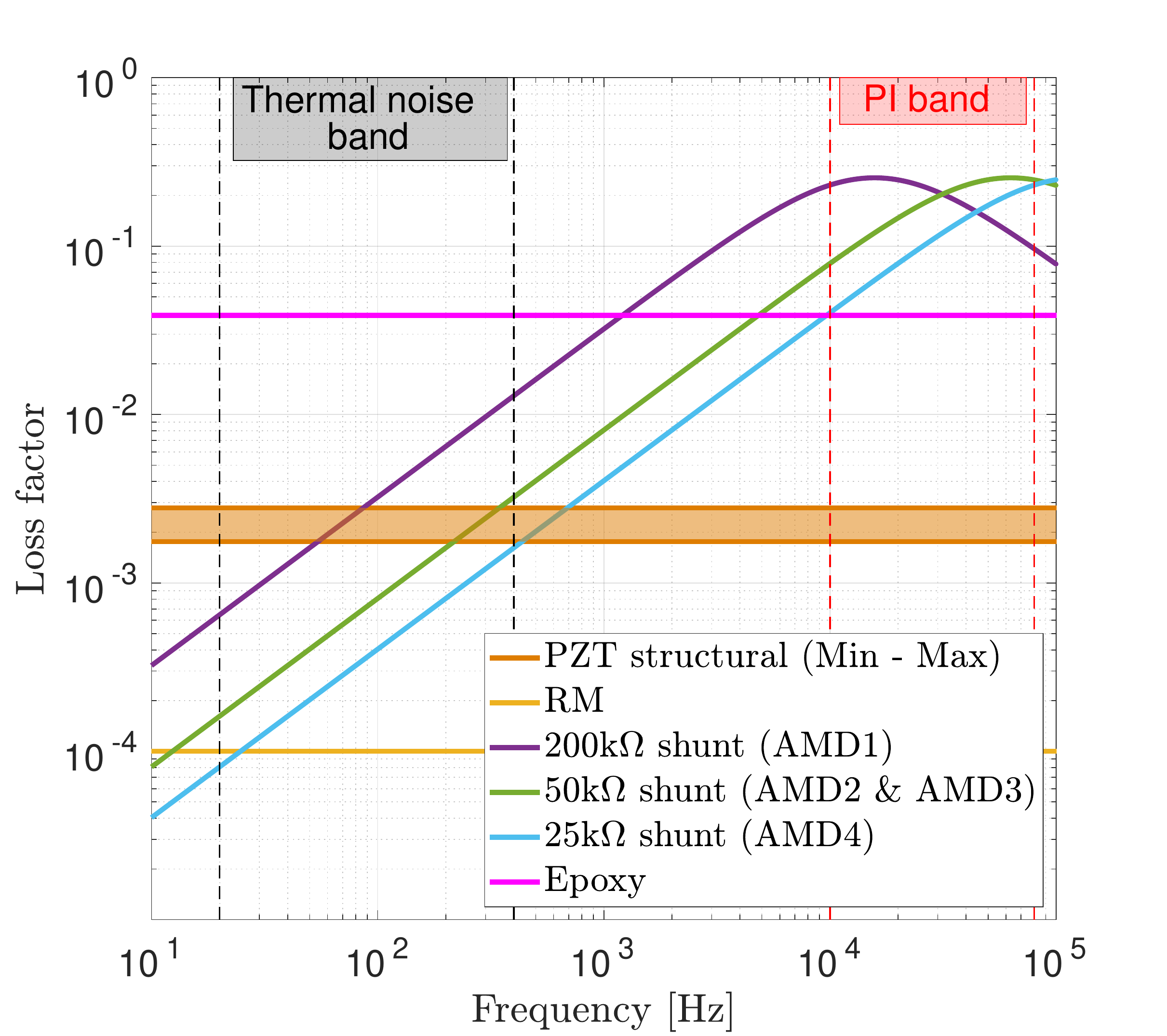}
	\caption{Loss factors of the different AMD materials as a function of frequency. The resistor values are chosen to maximize the damping efficiency of the AMDs active direction in the PI band, while limiting the thermal noise re-injection at lower frequencies.}
   \label{fig:loss_plot}
\end{figure}

The AMD concept is shown in Figure~\ref{fig:amd}. It consists of four key components: a base, a single piezoelectric plate (PZT) shunted with a resistor, 
a reaction mass and adhesive bonds used for the AMD assembly as well as for direct installation on the test masses. Each component is chosen carefully to 
limit its associated thermal noise. The components and their properties are summarized in Table~\ref{table:param} and Figure~\ref{fig:loss_plot}.

The main element of the resonator is a piezoelectric plate, which converts the strain energy of a mechanical mode into charge. 
This charge is shunted into a resistor to dissipate the electrical energy as heat. A shunted PZT is equivalent to a tunable lossy spring which, 
in conjunction with the reaction mass, determines the AMD principal resonances. In a one-dimensional model, corresponding to the PZT being loaded uniaxially
with either a normal or shear stress, the spring constant $K_{pzt,sh}$ of the shunted PZT is a function of the angular frequency $\omega = 2\pi f$: 
\begin{equation}
K_{pzt,sh}(\omega) = Y[1+i \eta_{r}(\omega)]\frac{S}{h} \,,
\label{eq:spring}
\end{equation}
where $Y$ is the Young's modulus of the PZT material (bulk or shear), $S$ is the surface area and $h$ the height of the plate.
The term $\eta_{r}$ is the loss due to the resistor shunting~\cite{hagood1991damping}:
\begin{equation}
\eta_{r}(\omega) = \frac{ RC\omega k^{2}}{(1-k^{2})+ (RC\omega)^{2}} ,
\label{eq:shunt}
\end{equation}
where $R$ is the shunt resistance and $C$ the capacitance of the PZT plate. The electromechanical coupling coefficient $k$ is a constant of the PZT material; 
its square represents the percentage of mechanical strain energy which is converted into electrical energy \cite{jaffe2012piezoelectric}. 
The peak value of $\eta_{r}$, which occurs at $\omega = \sqrt{1 - k^2} / RC$, is tuned with the resistor to the frequency range where most unstable modes exist. 
For this model we are neglecting mechanical loss in the PZT, but it will be included in the next section when calculating the thermal noise due to the AMD.

A reaction mass $m$ is attached to this lossy spring to create the AMD oscillator, with resonant frequency $f_D$, which is then attached to the test mass. The AMD and test mass system can be described
as a pair of coupled oscillators with a large mass ratio. The test mass acoustic mode we wish to damp is represented in this model by a mass $M$, equal to
the modal mass of the acoustic mode, attached to a fixed reference with a lossless spring, with a resonant frequency $f_m$. For this system of coupled oscillators,
reference~\cite{gras2015resonant} shows that the resulting $Q$-factor of the acoustic mode is:
\begin{equation}
    Q_m \simeq \frac{\eta_{r}^2 + (1 - \rho)^2}{\eta_r \mu \rho} \,,
\end{equation}
where $\rho = f_m/f_D$, and the mass ratio, $\mu = m/M$, is assumed to be small.

When the AMD resonance is near that of the test mass, $\eta_r \gg |1 - \rho|$ and the test mass mode $Q$-factor is reduced to $Q_a \simeq \eta_r / \mu$. We can thus estimate the 
size of the reaction mass required to reduce the $Q$-factors from $\gtrsim 10^7$ to $10^5$--$10^6$, sufficient to suppress PIs. With $\eta_r = 0.1$ and an
acoustic mode modal mass $M = 10\,$kg, this would require a reaction mass of 1--10$\,$mg. This simple model turns out to underestimate the required reaction mass, for a few reasons. One of these is that the AMD cannot always be placed at the point of maximum displacement of a given mode, which can be described as an effective increase of the modal mass by the square of the ratio of the displacement at the AMD location to that of the mode's antinode 
$M' = M(x_{\rm max}/x_{\rm AMD})^2$. Other factors include the multiple coupled degrees of freedom of the AMD and the directional nature of the piezo material, both of which are covered in the following section.

\subsection{Optimizing the AMD design}

Moving beyond this simple model, we need to include the loss of the PZT material and the loss of the adhesive used to bond the AMD elements to each other and to the test mass.
These loss factors are not significant for the acoustic mode damping, but they can be significant in the thermal noise band and therefore it is important to choose low-loss
materials. 

\begin{table*}[t]
\centering
\begin{threeparttable}
\begin{tabular}{p{3.25cm} p{3.75cm} p{4.0cm} p{2.0cm} p{3.5cm}}
\hline \hline
Component & Material & Dimensions & Mass & Loss factor \\ \hline \hline
Base                    & $\rm SiO_{2}$, Au               & $\phi\,5\,{\rm mm} \times 4\,{\rm mm}$   & 0.17$\,$g  & $ 1 \times 10^{-6}$\tnote{ a} \\
PZT, PI Ceramic        & PIC181, $\rm Pb(Zr,Ti)O_{3}$        & $3 \times 3 \times 1.5\,\rm mm^{3}$     & 0.11$\,$g  & $[1.76 - 2.79] \times 10^{-3}\,$\tnote{ b} \\ [0.6ex] 
RM1                      & Aluminum,                  & $\phi\,11.5\,{\rm mm} \times 2.0\,{\rm mm}$  & 0.53$\,$g  & $ 1 \times 10^{-4}$\tnote{ a} \\
RM2                      &    6061-T6,          & $\phi\,9.75\,{\rm mm} \times 1.5\,{\rm mm}$  & 0.27$\,$g  & $ 1 \times 10^{-4}$\tnote{ a} \\
RM3                      &  gold plated         & $\phi\,8.5\,{\rm mm} \times 1.0\,{\rm mm}$  & 0.12$\,$g  & $ 1 \times 10^{-4}$\tnote{ a} \\
RM4                      &          -         & $\phi\,5.5\,{\rm mm} \times 0.75\,{\rm mm}$  & 0.05$\,$g  & $ 1 \times 10^{-4}$\tnote{ a} \\ [0.6ex]
Resistor (shunt)        & $\rm TiO_{2}, Al_{2}O_{3}, epoxy$   & $2 \times 1.25 \times 0.55 \,\rm mm^{3}$  & 0.01$\,$g & $\rm 0.25\tnote{ c}$ \\
Epoxy, EPO-TEK          & 302-3M                  & $\rm 1.0 \,\mu m$ thick  & 13$\,\mu$g &  $\rm 38.8 \times 10^{-3}$\tnote{ b} \\
Epoxy (conductive)      & 302-3M+graphite                 & $\rm 1.2 \,\mu m$ thick  & 15$\,\mu$g &  $\rm 38.8 \times 10^{-3}$\tnote{ b} \\
\hline \hline
\end{tabular}
\begin{tablenotes}\footnotesize
\item[a] from ref. \cite{bansal2013handbook} and \cite{zener1940internal}
\item[b] measured using the test setup described in \cite{biscans2018method}
\item[c] $\eta$ peak values, see Fig.\ref{fig:loss_plot}.
\end{tablenotes}
\end{threeparttable}
\caption{List of AMD components and their properties. The reaction masses (RM1--4) are slightly different in size to target different frequencies, and their shape is non-circular to widen the effective bandwidth of each AMD. The loss factors were either extracted from literature or directly measured with the mechanical oscillator described in~\cite{biscans2018method}.}
\label{table:param}
\end{table*}

The thermal noise impact of the AMD can be further limited through careful choice of geometry. One of these choices takes advantage of the fact
that the test mass acoustic modes will generally exhibit surface displacement in all directions, while thermal noise is determined largely by
motion in the direction of the optic axis. Thus the PZT plate is mounted with its active direction perpendicular to the optic axis. 
Furthermore, a compressive PZT plate will always exhibit some charge generation even for accelerations orthogonal to the poling
directions due to bending of the plate. Therefore the AMD uses a shear plate PZT, to better isolate the active direction from
the optic axis direction.

Next we consider the size and shape of the reaction mass. Higher mass will provide more damping
of acoustic modes, but will also introduce more thermal noise. The latter can be understood 
qualitatively by considering that when the AMD experiences an acceleration, a higher reaction mass will induce more strain in the lossy elements of the AMD due to inertia. Thus we choose a reaction mass as small as possible, but still sufficient for acoustic mode damping. As shown in Table~\ref{table:param}, all of the reaction masses are less than 1$\,$g. In contrast
to the prototype presented in \cite{gras2015resonant}, the reaction mass is made from a low-density 
material (aluminum) so that its moment of inertia can be increased without increasing its mass. This means we can achieve the desired mechanical resonant frequencies of the AMD assembly using less mass, thereby limiting the thermal noise impact.
Finally, the reaction mass shape is intentionally not symmetric (see Fig.~\ref{fig:amd}), which breaks the degeneracy of principal resonances in orthogonal directions to widen the effective
bandwidth of a single AMD.

The size of the fused silica base is also chosen to minimize thermal noise. To do this it is important to minimize the area of the bond to the test mass, so the base diameter is just large enough for the PZT. The base height of 4$\,$mm is larger than it needs to be so that in the thermal noise band, the AMD structure deforms mostly in the low-loss base, rather than in the higher-loss PZT.
 
Finally, epoxies are used to bond the AMD elements together and to bond the AMD to the test mass. Though the volume and mass of the epoxies are much smaller than that of the other elements, epoxies have relatively high mechanical loss and they need to be chosen carefully. 
Several epoxies were evaluated in terms of their minimum bond thickness, curing requirements, and mechanical loss. The loss factor of the chosen epoxy (see Table~\ref{table:param}) 
displayed a significant dependence on bond thickness~\cite{biscans2018optimization}, becoming larger for thicknesses less than a couple of microns. Thus the thermal noise impact of the
epoxy is not minimized by making the thinnest possible bond; instead we found the optimal bond thickness to be approximately 1 micron. The bonds to the PZT plate require a conductive medium, 
and for these we mixed graphite nano-powder with the epoxy. We confirmed that the graphite-filled epoxy had the same loss factor as the regular epoxy. 


\subsection{Modal damping efficiency}
\label{sec:perf}

\begin{figure}[h!]
	\centering
	\includegraphics[width=0.45\textwidth]{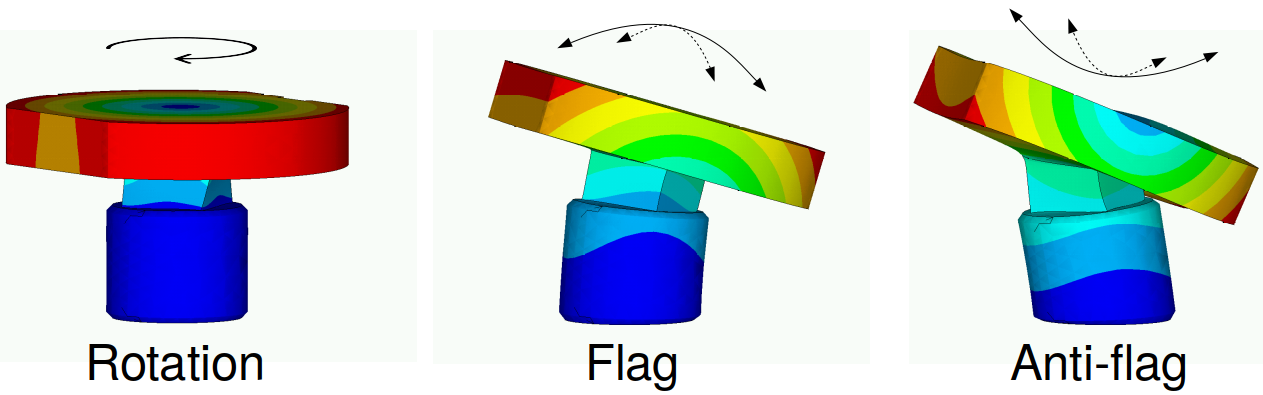}
	\caption{Three different types of AMD principal resonances which have non-zero strain in the active PZT direction (shear). There are five principle resonances in total per AMD. Due to the asymmetry of the reaction mass and the 45 degree orientation of the PZT plate, the flagpole and anti-flagpole modes appear in doublets. The torsion mode is also effective for damping due to the anisotropy of the PZT material. }
   \label{fig:amd_modes}
\end{figure}

Efficient damping of the test mass acoustic modes requires that the AMD principal resonances have 
good overlap in frequency with these modes. The AMD design has five principal resonances: two bending or `flagpole' resonances, two anti-flagpole resonances, and a single torsional mode, as shown in Fig.~\ref{fig:amd_modes}. Each of these modes involves large strain in the active direction of the PZT element, and efficient conversion of mechanical energy to electrical energy. The compression mode is not considered herein, as it has very little coupling to shear in the PZT plates.

Models of a test mass with various numbers of AMDs attached were analyzed via finite element analysis (FEA)~\cite{ansys}. The test mass is a right circular cylinder ($34\,{\rm cm}\, \phi \times 20\,{\rm cm}$ thick), with two flats polished on opposing sides (see Fig.~\ref{fig:amd}). For ease of attachment, the AMDs are mounted on these flats, within a specific area at the top of the flat and adjacent to the test mass front face. The test mass modes are first calculated with FEA in the absence of AMDs. This model uses a bulk loss for the fused silica of $10^{-7}$, and includes the much higher loss ($\simeq 10^{-4}$), several-micron thick coating that creates the mirror surface. The acoustic mode $Q$-factors from this model ({\it sans} AMD) range from 10-40 million for modes in the 15-80$\,$kHz band.

Modeling the system with AMDs mounted on the flats, we found that a set of at least four AMDs with evenly spaced principal resonances is required to cover the entire PI frequency band. The AMD resonances are spread out by using different RM dimensions and masses for each AMD, and different shunting resistors are used  
to spread the peaks of $\eta_{r}$ across the (15-80)$\,$kHz range.
The properties of each AMD are given in Table~\ref{table:param}.

\begin{table}[b!]
\centering
\begin{tabular}{ l   c }
\hline
\hline 
Input test mass RoC                     &  (1936 - 1945) m  \\ 
End test mass RoC                     &  (2248 - 2254) m  \\
Acoustic mode $f_m$ uncertainty   & $\pm$ 2\%        \\ 
\hline
SRC Gouy phase              & 19 deg.           \\ 
PRC Gouy phase                    & 25 deg.           \\ 
No. of mechanical modes      &  4,200           \\ 
No. of iterations            & 200,000             \\\hline \hline
\end{tabular}
\caption{Monte Carlo parameters for computation of the expected parametric gain, with the varied parameters listed in the first 3 rows. One-way Gouy phases for the signal recycling and power recycling cavities (SRC and PRC, respectively) are held constant, whereas the radii-of-curvature (RoC) of the test masses and the acoustic mode eigenfrequencies are varied.} 
\label{table:MC}
\end{table}

The quality factor $Q_{m}$ for a test mass mechanical mode with AMDs attached is calculated with the following formula:
\begin{equation}
Q_{m}(f_{m}) = \frac{ E_{s}(f_{m})}{\sum\limits_{i} E_{i}(f_{m}) \eta_{i}(f_{m})} \,,
\label{eq:tn}
\end{equation}
where $E_{s}$ is the total modal strain energy of the test mass+AMD, and $E_{i}$ is the strain energy of the individual component $i$ with the loss factor $\eta_{i}$. The sum is over all of the AMD elements listed in Table~\ref{table:param}, as well as the test mass elements that are in the model (bulk and coating). The strain energy values are obtained with the FEA, and the loss values are taken from Table~\ref{table:param} and Fig.~\ref{fig:loss_plot} (for the $\eta_r$ values). For the set of four AMDs, 98\% of the acoustic mode $Q$s are suppressed by a factor of 10 or more compared to their values without AMDs, and if $f_m$ is very close to an AMD principal resonance, the suppression factor can be 100 or more. 

\begin{figure}[h!]
	\centering
	\includegraphics[width=0.5\textwidth]{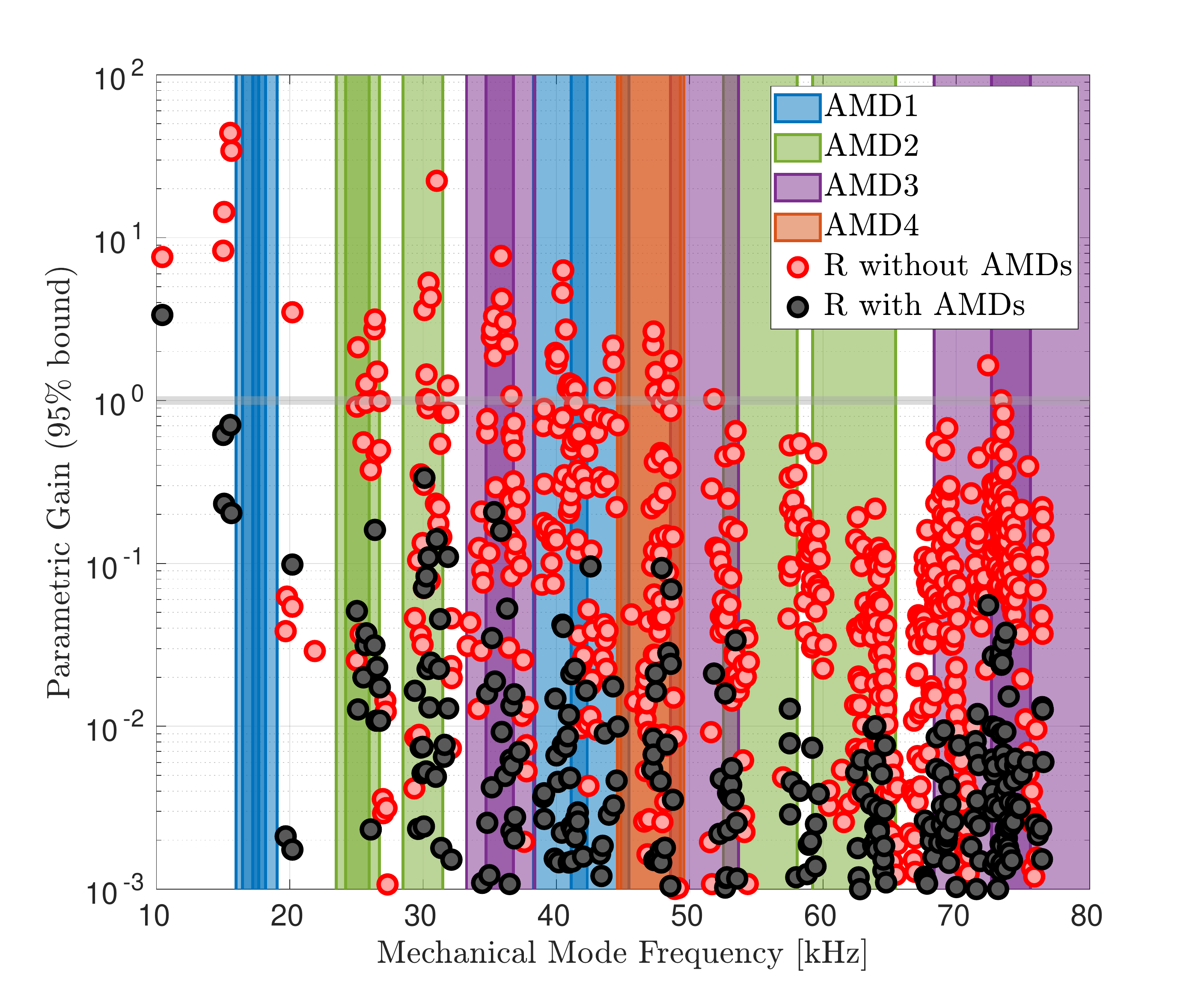}
	\caption{Comparison of the expected parametric gains at full power ($P_{arm} = 750\,$kW) without AMDs (\textit{red circles}) and with AMDs (\textit{black circles}), for a single test mass. Each data point corresponds to the 95\% bound on the gain, $R_{95}$, as
	explained in the text. With AMDs, all modes at and above 15$\,$kHz should become stable ($R < 1$). Each colored vertical bar corresponds to a principal mode of that AMD, with the bar width indicating the resonance 3~dB points. Four AMDs provide good overlap of AMD principal resonances with all potentially unstable mechanical modes above 15$\,$kHz.}
   \label{fig:perf2}
\end{figure}

These $Q_m$ values from the FEA can be used to calculate the parametric gain $R_m$ using Eq.~\ref{eq:rm}, but uncertainties in several of the parameters prevent an accurate calculation of the gain for a given mode. Instead, we use the Monte Carlo method described in \cite{evans2010general} to determine the range of potential parametric gain values for each acoustic mode.
The key parameters for this simulation are given in Table~\ref{table:MC}. 
Each arm cavity comprises a partially-transmissive `input test mass', and a highly-reflective `end test mass', which
differ only in their mirror coatings and radii-of-curvature. The FEA parameters (mode frequencies and $Q$s) for an
end test mass are used in this PI analysis.
From the Monte Carlo results, we identify the 95\% bound on the parametric gain$\,$--$\,$i.e., the level that
95\% of the values do not exceed$\,$--$\,$and denote this value as $R_{95}$.
The results for the target design power in the arm cavities (750$\,$kW) are given in Fig.~\ref{fig:perf2}, which shows that all modes between (15-80)$\,$kHz should be stable when the test masses are outfitted with AMDs. For the mode at 15.5$\,$kHz, which is the strongest PI observed in LIGO, $R_{95}$ is reduced from 44 down to 0.7. 

One mechanical mode at 10.4$\,$kHz is still likely to produce an instability at full power,  with an $R_{95}$ of 3.4. This is a drum head mode of vibration, where there are no nodal diameters and the faces of the test mass vibrate primarily along the cavity optic axis (similar to the fundamental mode of a circular membrane). Since this mode shape is similar to the test mass deformation relevant for thermal noise, the AMDs are designed to avoid coupling to it to minimize their thermal noise impact. Furthermore, the mode has an extremely high $Q$-factor; the FEA predicts an intrinsic $Q$ of 62 million, which is damped only to 30-40 million by the AMDs. This mechanical mode couples mainly to a second order transverse optical mode, the Laguerre-Gauss $LG_{1,0}$ cavity mode.  
The instability associated with this mode can still be controlled via thermal tuning, and should not present a limitation.

\subsection{Thermal noise estimation}
\label{sec:TN}

The power spectral density $S_t$ of thermal noise fluctuations can be computed using the generalized fluctuation-dissipation theorem~\cite{callen1951irreversibility}. We follow Levin's method~\cite{levin1998internal} and use FEA harmonic analysis to compute $S_t$:
\begin{equation}
S_t(f) = \frac{4k_{B}T}{\pi f F_{0}^{2}}  \sum\limits_{i} E_{i}(f)  \eta_{i}(f) \,, 
\label{eq:thermal_noise}
\end{equation}
where $k_{B}$ is Boltzmann's constant and $F_{0}$ is the amplitude of an oscillating pressure field applied to the front surface of the test mass model. The spatial profile of the pressure field corresponds to
the laser beam intensity incident on the test mass$\,$--$\,$a fundamental mode Gaussian with a beam radius of either 6.2$\,$cm (end test mass) or 5.3$\,$cm (input test mass). The pressure field creates a deformation in each element of the model, and from the
FEA we can extract the strain energy $E_{i}$ in each element. The total thermal noise due to each
AMD is found by summing Eq.~\ref{eq:thermal_noise} over its elements, with loss factors coming from 
either Fig.~\ref{fig:loss_plot} (for $\eta_r(\omega)$) or Table~\ref{table:param} (all other elements).

The FEA thermal noise results are shown in Table~\ref{table:TN_results}. For one test mass,
the estimated thermal noise from four AMDs is $\rm 1.16 \times 10^{-21} m/\sqrt{Hz}$ at 100$\,$Hz. With all four interferometer 
test masses (16 AMDs), this corresponds to a total noise contribution of $\rm 2.32 \times 10^{-21} m/\sqrt{Hz}$ at 100$\,$Hz. 
This is to be compared to the target design sensitivity of 
Advanced LIGO~\cite{aasi2015advanced} at 100$\,$Hz, which, at $\rm 16.3 \times 10^{-21} m/\sqrt{Hz}$, is
dominated by quantum noise and thermal noise from the test mass mirror coatings.

\begin{figure}[h!]
\centering
\includegraphics[width=0.45\textwidth]{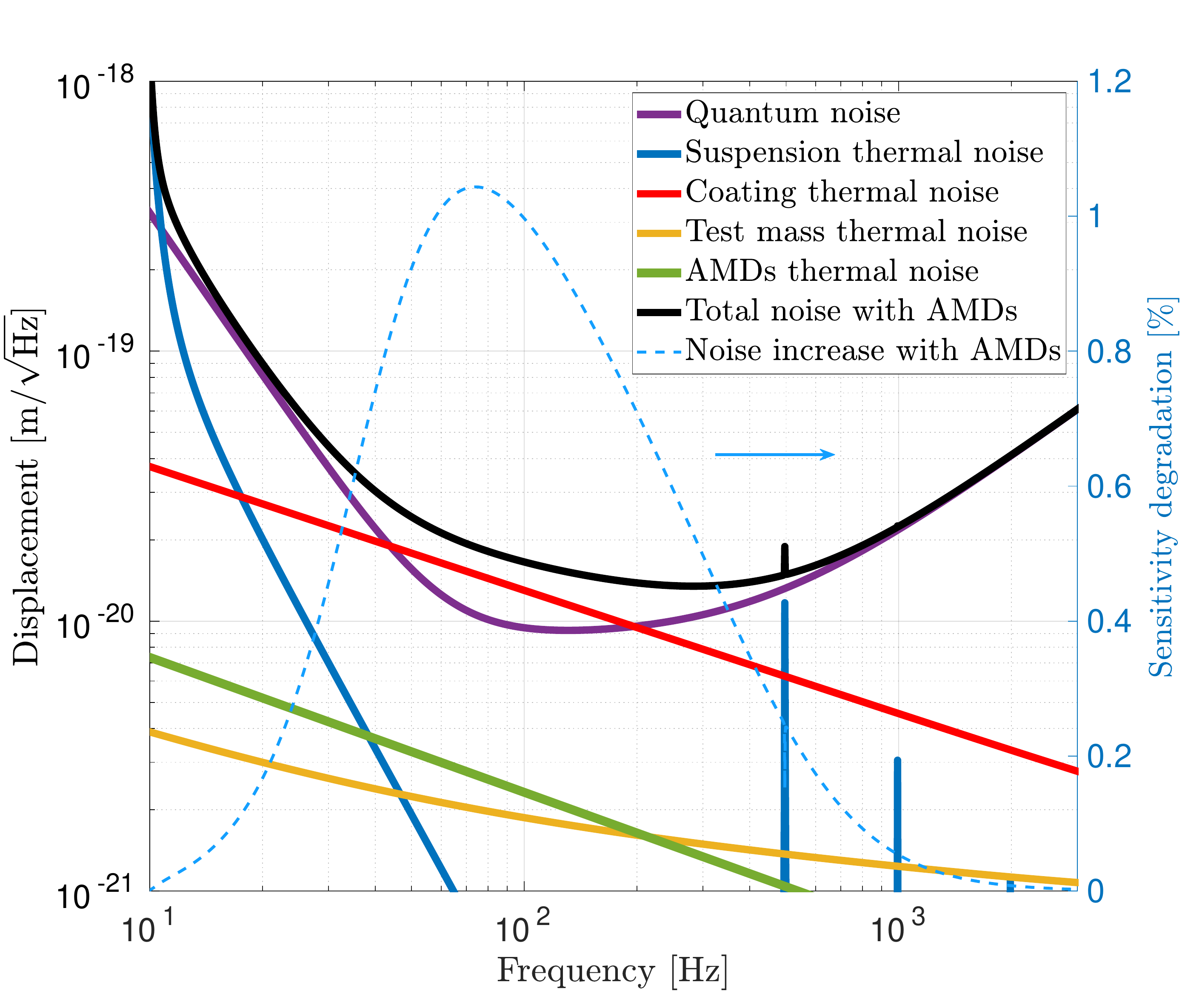}
\caption{Displacement noise amplitude spectra for AMD thermal noise (red curve) and the major noise contributors to the Advanced LIGO design at full power, $P_{arm} = 750\,$kW. Equivalent detector strain
noise is derived by dividing by 4000$\,$m. The AMD curve corresponds to 16 total AMDs (4 per test mass). The blue dashed line shows the sensitivity degradation in percent as a result of adding the AMDs.}
\label{fig:TN_gwinc}
\end{figure}

The spectrum of displacement thermal noise due to the AMDs is shown in Fig.~\ref{fig:TN_gwinc}, along
with the Advanced LIGO design spectrum and its noise contributors. The plot also shows the degradation
of the design spectrum due to the AMDs, indicating a maximum noise penalty of 1.0\% at 70$\,$Hz.

\begin{table}[t!]
\centering
\begin{tabular}{| c | c | c | c | c |}
\hline
       &  \multicolumn{4}{c|}{Thermal noise at 100$\,$Hz}  \\
       &  \multicolumn{4}{c|}{[$\rm \times 10^{-22} m/\sqrt{Hz}$]}         \\ \hline \hline
                  & \textbf{ AMD1} & \textbf{AMD2} & \textbf{AMD3} & \textbf{AMD4}  \\ \hline
Base              & 0.19 & 0.12 & 0.07 & 0.06  \\ \hline
RM  & 0.50 & 0.23 & 0.10 & 0.03  \\ \hline
\textit{Epoxy between:}    &      &      &      &                               \\ 
Test mass \& Base        & 6.52 & 4.19 & 3.00 & 2.46 \\ 
Base \& PZT       & 4.16 & 2.24 & 1.21 & 0.73  \\ 
PZT \& RM         & 2.49 & 1.23 & 0.54 & 0.2               \\ \hline
PZT (structural)    & 4.48 & 2.31 & 1.16 & 0.62  \\ \hline
PZT (shunt)    & 0.031 & 0.009 & 0.011 & 0.003  \\ \hline
\textbf{Total AMD} & \textbf{9.30} & \textbf{5.43} & \textbf{3.48} & \textbf{2.65}  \\ \hline \hline
\multicolumn{5}{|c|}{\textbf{Total noise for 1 test mass}  $\rightarrow$ \textbf{11.62}}  \\ \hline
\end{tabular}
\caption{Thermal noise budget of the four AMDs at 100$\,$Hz for 293$\,$K. The thermal noise level is strongly correlated with the mass of the reaction mass. The largest thermal noise contributors are the epoxy layers and the PZT material. The very small contribution from the shunt is a result of orienting the PZT polarization direction perpendicular to the cavity optic axis.} 
\label{table:TN_results}
\end{table}

\section{Experimental results - PI mitigation}

For LIGO's O3 observing run, all test masses at both observatories have been fitted with the
set of four AMDs described above. No parametric instabilities are observed in the (15-80)$\,$kHz range, even 
without implementing any thermal mode tuning or active damping. This is at an arm power level of $P_{arm} = 230\,$kW,
in contrast to the situation without AMDs, when the first instability would appear at 25$\,$kW arm power.

To quantitatively assess the performance of the AMDs, we made three types of measurements: $Q$-factor measurements of test mass acoustic modes, with and without AMDs; parametric gain of a specific mode versus thermal cavity geometry tuning; and a noise measurement to bound the thermal noise impact. 

\subsection{Test Mass $Q$-factors}

Suspended adjacent to each test mass is a reaction mass that includes a pattern of electrodes which can be driven to apply electro-static forces to the test mass.
These actuators are used to excite the test mass acoustic modes and measure their $Q$-factors from the ring-downs recorded in the main gravitational-wave channel.
Test mass modes were excited while the interferometer was operating at low laser power, in order to avoid parametric gain significantly altering the ring-down times.

\begin{figure}[t!]
\centering
\includegraphics[width=0.5\textwidth]{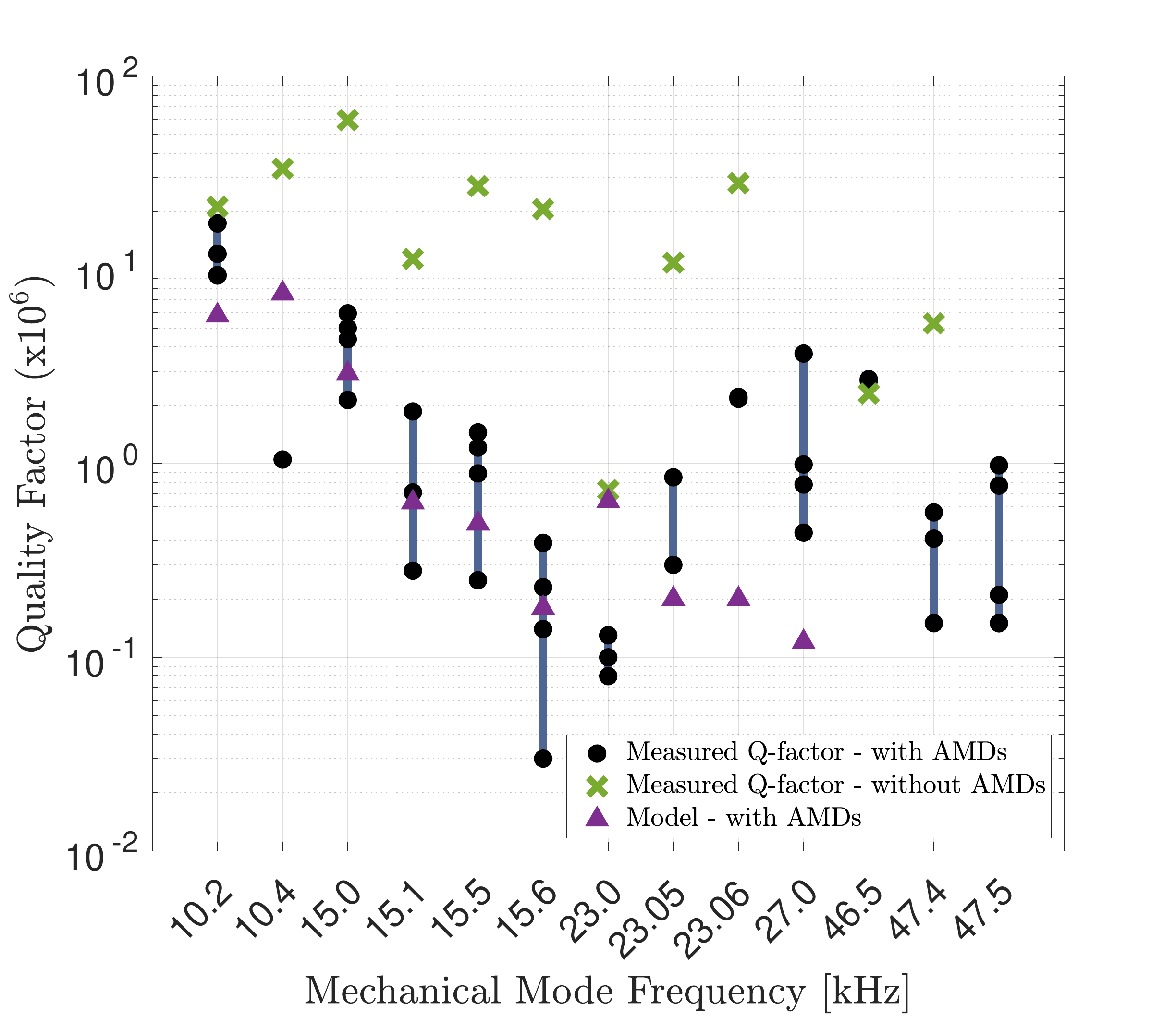}
\caption{Measured $Q$-factors of test mass acoustic modes. The green crosses correspond to pre-AMD measurements of one of the Livingston Observatory test masses. The post-AMD $Q$-factors are shown as black dots. The blue bars indicate the spread of $Q$s measured across several test masses (with AMDs), and the purple triangles represent the model prediction of $Q$s with AMDs.}
\label{fig:Q_meas}
\end{figure}

We were able to measure $Q$-factors for thirteen modes, usually on multiple test masses, in the band ($10-50)\,$kHz. For the ten lowest frequency modes, we could
identify their particular mode shapes and so can compare the measurements to the finite element model predictions (above 30$\,$kHz the mode density
is so high that it is not possible to uniquely identify the modes). The $Q$-factor measurements from the LIGO Livingston interferometer are shown in Fig.~\ref{fig:Q_meas};
the results from the Hanford interferometer are similar. The plot also includes eleven $Q$-factors from one of the Livingston test masses measured before the AMDs
were installed. 
As expected, the $Q$-factors for all but one of the modes at 15$\,$kHz and above are reduced by nearly an order of magnitude or more.
The variations in $Q$ from test mass to test mass and from the modelled values are not too surprising given realistic deviations in AMD and test mass parameters; for example, 
any frequency mismatch between the AMD resonances and the test mass modes will reduce the damping. The FEA of
the test mass predicts acoustic mode frequencies with a typical error of 0.5\%, or up to a few hundred Hz. In addition, some AMD parameters are difficult 
to control during assembly and installation. We estimate that the epoxy bond thickness could vary by up to -50\% or +20\% from the 1$\,\mu$m nominal thickness, which would limit the accuracy of the AMD principal resonance to about 5$\,$kHz. Also, the installed locations of the AMDs on the test masses could differ from
the model by
several mm, due to varying mounting constraints from test mass to test mass.

\subsection{Optical mode transient test}

\begin{figure}[t!]
\centering
\includegraphics[width=0.51\textwidth]{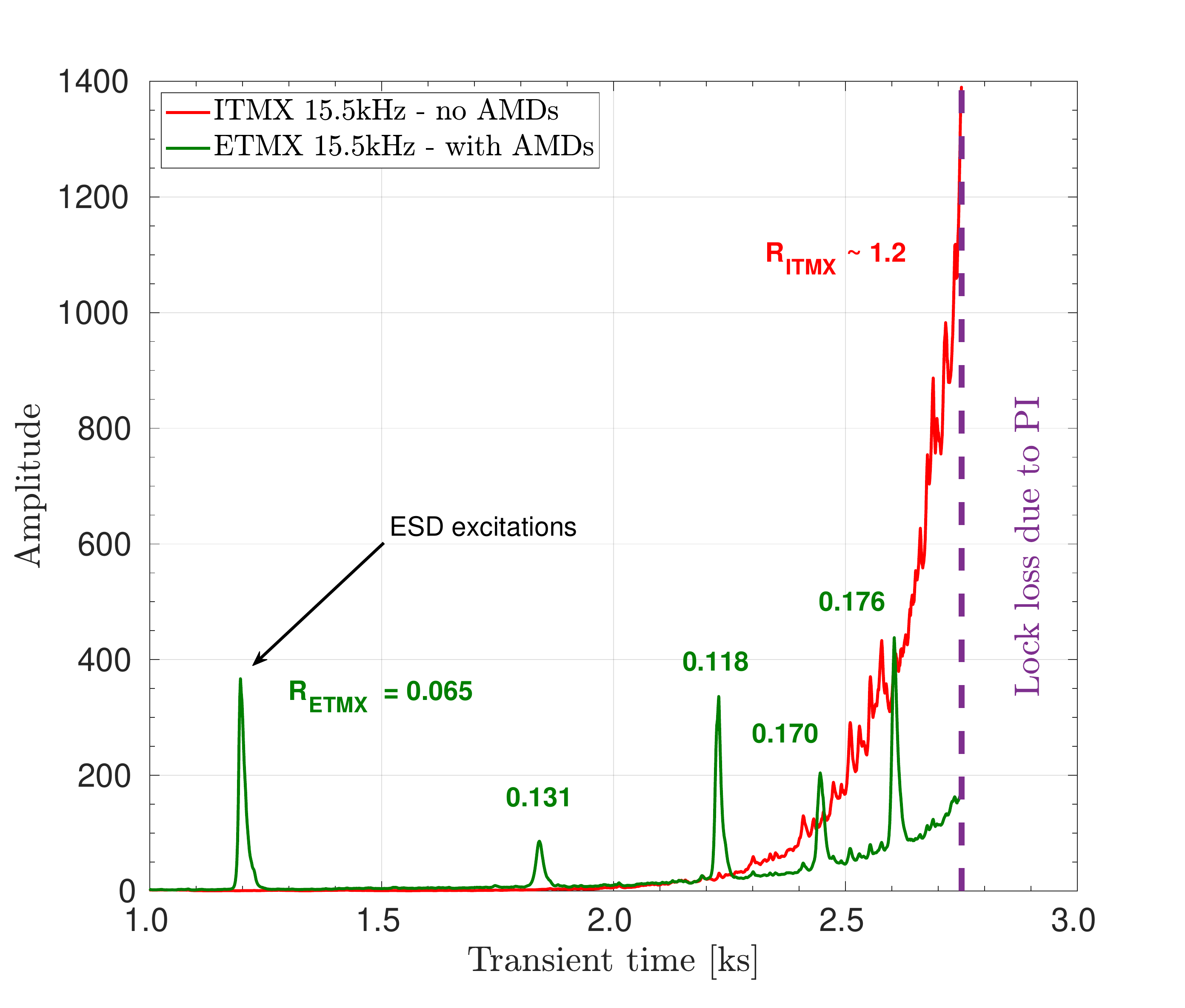}
\caption{Thermal transient of the 15.5kHz modes on input (ITMX) and end (ETMX) test masses, respectively. As expected, the ITMX without AMDs became unstable during thermal transient, with rapidly rising amplitude. Contrary, ETMX which has attached AMDs remains stable with parametric gain below unity. The rising envelope of the ETMX signal is a result of imperfect filtering out of the ITMX signal.}
\label{fig:ringdown}
\end{figure}

According to Eqn.~\ref{eq:rm}, parametric gain scales linearly with the gain of the higher-order optical mode $G_n$, assuming one relevant optical mode. $G_n$ in turn depends on how close the
acoustic mode frequency is to the optical mode frequency~\cite{braginsky2001parametric}: $G_n \propto (\Delta f_n^2 + 4 \Delta f^2)^{-1}$, 
where $\Delta f_n$ is the linewidth of the higher-order mode $n$, and $\Delta f = f_m - f_n$, where $f_n$ is the frequency of the higher-order mode relative
to the frequency of the arm cavity ${\rm TEM_{00}}$ mode.
Thermal tuning of a test mass' curvature will shift $f_n$, and thereby change $\Delta f$ and the optical gain. By tuning an optical mode to be very
close to its acoustic mode PI-partner, $f_n \approx f_m$, we can determine the maximum possible parametric gain for that acoustic mode.

We performed such a measurement when AMDs were installed on a single end test mass in one arm cavity (the X-arm) at the LIGO Livingston Observatory. We locked the interferometer
with 100$\,$kW of power in the arm cavities, and took advantage of the small absorption in the mirror coatings (sub-ppm) which creates a thermal tuning
transient with a time constant of approximately 1 hour. 
We monitored the amplitude of the 15.5$\,$kHz acoustic mode of both X-arm test masses, the mode most prone to instability through interaction with a third-order 
 transverse optical mode; details of these mechanical and optical modes can be found in ~\cite{evans2015observation}. 
The transient thermal tuning shifts the third-order optical mode higher in frequency, towards 15.5$\,$kHz.  The acoustic mode is separated by about 4$\,$Hz between the two test masses (one with AMDs and one without), so the
light scattered from each experiences nearly the same optical gain as $\Delta f$ changes.  We periodically excited the 15.5$\,$kHz mode of the end test mass with its
electro-static actuator and measured the ring-down time, from which the parametric gain was extracted using Eq.~\ref{eq:tau}.

The evolution of the 15.5$\,$kHz mode amplitude in both test masses is shown in  Fig.~\ref{fig:ringdown}. The input test mass, which did not have AMDs, becomes unstable with a
measured parametric gain of $R = 1.2$ before it drives the interferometer out of lock. On the other hand, the end test mass, with AMDs, remained stable with a highest measured gain of $R = 0.176$. 
While we cannot be certain that this corresponds to the highest possible optical gain, Fig.~\ref{fig:ringdown} shows
that $R \leq 0.176$ for a range of thermal tunings. 
Since the 15.5$\,$kHz mode is the strongest in terms of parametric instabilities, this end test mass $R$ value can be used to estimate the maximum arm power at which the interferometers should be stable under most thermal tuning conditions: $P_{max} = 100\,{\rm kW}/0.176 = 570\,$kW.  Furthermore, any instabilities that occur when the full design power of 750$\,$kW is reached should be avoidable with thermal cavity tuning.

\subsection{Thermal noise impact}

The additional thermal noise introduced by the AMDs is expected to be small, increasing the detector's design strain noise by at most 1.0\% at 70$\,$Hz.
It is not possible to verify the thermal noise impact at that level, but we can set an upper limit by comparing the measured interferometer noise to noise model
expectations, and to measured noise before AMDs were installed. 
An increase in thermal noise would first be evident in the band (40-200)$\,$Hz (see Fig.~\ref{fig:TN_gwinc}), but the detector's noise spectrum is limited by quantum shot noise at frequencies above 50$\,$Hz range, masking thermal and other classical noises. The classical noise spectrum underneath the quantum noise can, however,
be revealed using the cross-correlation technique described in~\cite{martynov2017quantum}. This technique takes advantage of the fact that the light at the
output port is split into two equal intensity beams, and homodyne detection is performed on each beam. Quantum shot noise and photodetector dark noise are
uncorrelated in these two detection channels, and therefore their contribution to the cross-spectrum of the two channels diminishes with more averages, leaving the coherent, classical noise.

Data was analyzed for the Livingston detector during low-noise operating states both before and after all AMDs were installed. 
Between O2 and O3 several detector improvements and changes were made in addition to the AMDs, so the before/after comparison of classical noise does not test only the effect of
the AMDs. However, it can be used to verify that the classical noise did not increase with the presence of AMDs. In addition, we can compare the measured
cross-spectrum with the modeled classical noises, which are well-known in the frequency band of interest.

\begin{figure}[t!]
\centering
\includegraphics[width=0.5\textwidth]{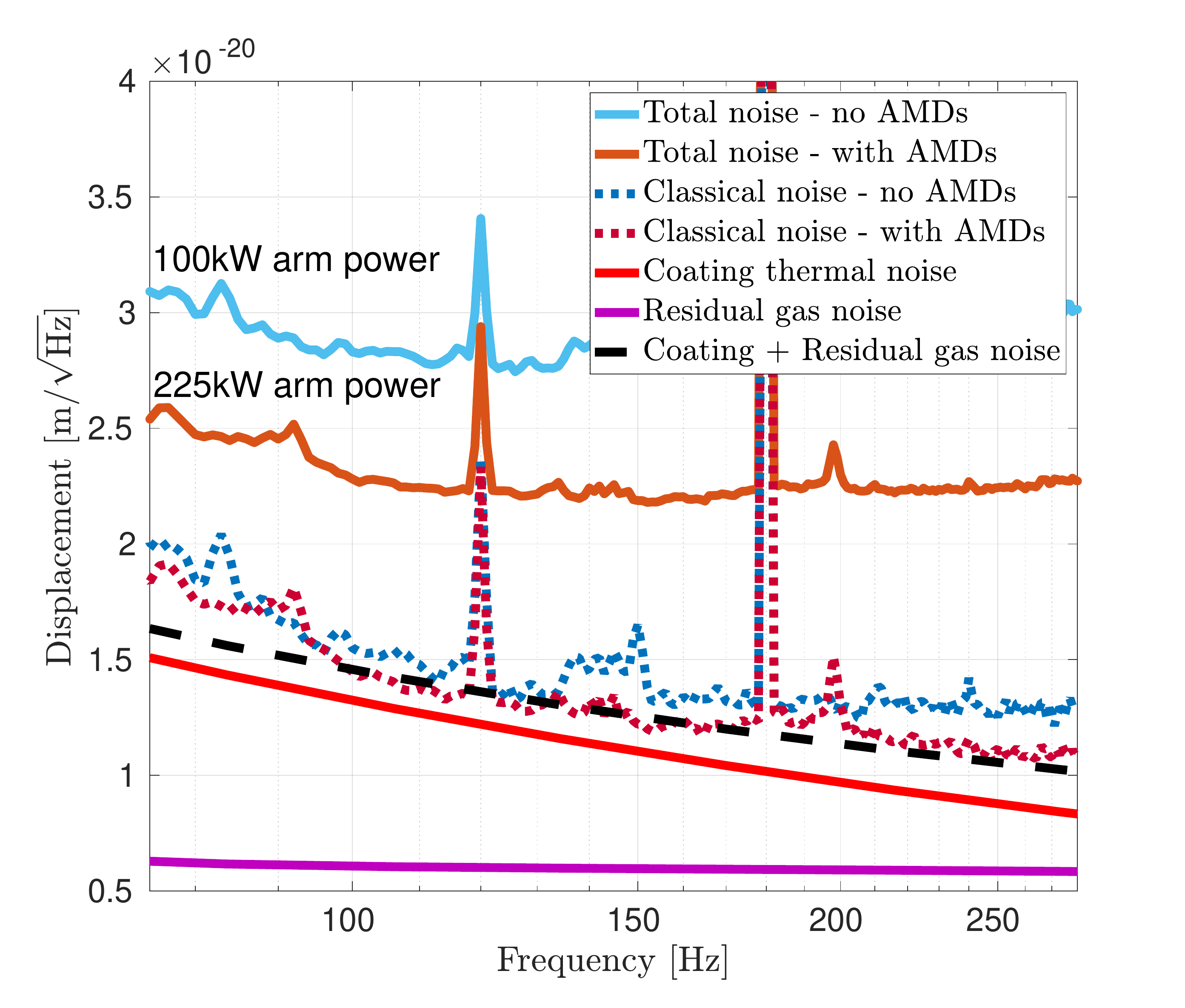}
\caption{Noise spectra of the Livingston interferometer before and after installation of all AMD. The solid lines show the total noise level measured (classical + quantum noise). The dotted show the level of classical noise only, after the quantum has been subtracted via a cross-correlation technique. Coating thermal noise and residual gas noise - the dominant classical noise contributions in this region - are also shown.}
\label{fig:cross_corr_AMD}
\end{figure}

Fig.~\ref{fig:cross_corr_AMD} shows spectra of the total interferometer noise and the classical noise measured with the cross-spectrum, both with and without
AMDs on the test masses. The total noise in the case with AMDs is lower than the earlier, {\it sans} AMD data due to higher circulating arm power (225$\,$kW vs. 100$\,$kW),
which reduces the quantum shot noise contribution. The small decrease in classical noise after AMD installation is likely due to an unrelated reduction in a different classical noise, such as scattered light. We see that the measured classical noise with AMDs (dashed red curve) matches the noise model estimate for the dominant classical noises well in most of this frequency band.  The discrepancy between the AMD measurement 100-150$\,$Hz and the  model is within the $\sim2\%$ detector calibration uncertainty~\cite{ligoc2017alibration} and model uncertainties which are larger than calibration uncertainty.
These model contributions are coating thermal noise and phase noise due to residual gas in the beam tubes. 
There is no evidence that the AMDs are introducing significant additional thermal noise.

\subsection{Effect of beam decentering}

The PI simulations that produced the data in Fig.~\ref{fig:perf2} assumed the cavity
beams are centered on the test mass faces. In practice, during the O3 observing run the cavity beams are intentionally decentered on several of the test masses in order to avoid small defects in their mirror coatings. The typical beam decentering of 20$\,$mm can significantly increase the geometrical overlap $B_{m,n}$ between some mechanical and optical modes, thereby increasing their PI probability. This is
particularly the case for a pair of acoustic modes at 10.2 and 10.4$\,$kHz, which have a displacement pattern on the test mass face similar to the Zernike trefoil polynomial. These
modes overlap only weakly with the Hermite-Gaussian second order modes 
($HG_{0,2}, HG_{2,0}$ and $HG_{1,1}$)
when the cavity beam is centered, but the overlap factor can increase by several orders
of magnitude when the beam is off-center. For example, $B^2_{m,n}$ between the $HG_{1,1}$
mode and the 10.4$\,$kHz mode increases from $2 \cdot 10^{-8}$ to $6 \cdot 10^{-3}$ for
a decentering of 18$\,$mm. 

Indeed, instabilities have been observed at both 10.2 and 10.4$\,$kHz in one arm of the LIGO 
Hanford interferometer, at a power level of $P_{arm} = 230\,$kW. The 10.4$\,$kHz mode could
be either the trefoil mode just mentioned, or the drumhead mode mentioned in Sec.~\ref{sec:perf}; it is difficult to distinguish between the two as their eigenfrequencies 
differ by only $\sim\!10\,$Hz. Both instabilities are stabilized by shifting the second-order optical modes by $\sim\!100\,$Hz using the ring heater on the end test mass (i.e., thermal cavity geometry tuning).

\section{Conclusion}

We have presented a simple yet effective passive device to mitigate parametric instabilities in interferometric gravitational-wave detectors. 
The significant advantage of these acoustic mode dampers compared to previous mitigation techniques~\cite{zhao2005parametric, blair2017first} is that they act on all instabilities simultaneously without requiring further tuning or intervention. 
Acoustic mode dampers designed to provide tuned damping of the (15-80)$\,$kHz internal modes of the LIGO test masses have been installed on all four 
test masses in both LIGO interferometers. With these dampers in place, no instabilities have
been observed in the (15-80)$\,$kHz range at arm circulating powers as high as 240$\,$kW. 
Two instabilities have been observed in one interferometer near 10$\,$kHz, and these have
been controlled with a small amount of thermal tuning. Importantly, no active damping has been 
required on either interferometer to achieve long-term stability.

At the full design power of $P_{arm} = 750\,$kW, assuming the cavity beams are then centered
on the test masses, the AMDs should stabilize all acoustic modes except for the 10.4$\,$kHz
drumhead mode, and possibly the 15.5$\,$kHz modes. The drumhead mode will need to be
stabilized, most likely with thermal tuning (active damping may be difficult as the 
electro-static actuators do not couple strongly to this mode). If we can improve the accuracy with which the AMD principal resonances can be made to match their design values, we can
better target the 15.5$\,$kHz modes to decrease their probability of being unstable.

Attaching any components to the test masses must be done carefully in order to avoid increasing the thermal noise in the gravitational-wave 
detection band. Our model predicts that the detector's equivalent strain noise will be degraded by at most 1.0\% around 70$\,$Hz by the
addition of the AMDs. Our measurement of the classical noise present in the interferometer is not accurate enough to verify such a small
impact, but it does show that there is no significant increase in thermal noise.


\begin{acknowledgments}

The authors acknowledge the entire LIGO Scientific Collaboration for their wide ranging expertise and contributions. LIGO was constructed by the California Institute of Technology and Massachusetts Institute of Technology with funding from the National Science Foundation, and it operates under Cooperative Agreement No. PHY-1764464. Advanced LIGO was built under Grant No. PHY-0823459. 
This paper carries LIGO Document Number LIGO-P1900243.
\end{acknowledgments}

\bibliography{ref} 

\begin{thebibliography}{29}%
\makeatletter
\providecommand \@ifxundefined [1]{%
 \@ifx{#1\undefined}
}%
\providecommand \@ifnum [1]{%
 \ifnum #1\expandafter \@firstoftwo
 \else \expandafter \@secondoftwo
 \fi
}%
\providecommand \@ifx [1]{%
 \ifx #1\expandafter \@firstoftwo
 \else \expandafter \@secondoftwo
 \fi
}%
\providecommand \natexlab [1]{#1}%
\providecommand \enquote  [1]{``#1''}%
\providecommand \bibnamefont  [1]{#1}%
\providecommand \bibfnamefont [1]{#1}%
\providecommand \citenamefont [1]{#1}%
\providecommand \href@noop [0]{\@secondoftwo}%
\providecommand \href [0]{\begingroup \@sanitize@url \@href}%
\providecommand \@href[1]{\@@startlink{#1}\@@href}%
\providecommand \@@href[1]{\endgroup#1\@@endlink}%
\providecommand \@sanitize@url [0]{\catcode `\\12\catcode `\$12\catcode
  `\&12\catcode `\#12\catcode `\^12\catcode `\_12\catcode `\%12\relax}%
\providecommand \@@startlink[1]{}%
\providecommand \@@endlink[0]{}%
\providecommand \url  [0]{\begingroup\@sanitize@url \@url }%
\providecommand \@url [1]{\endgroup\@href {#1}{\urlprefix }}%
\providecommand \urlprefix  [0]{URL }%
\providecommand \Eprint [0]{\href }%
\providecommand \doibase [0]{http://dx.doi.org/}%
\providecommand \selectlanguage [0]{\@gobble}%
\providecommand \bibinfo  [0]{\@secondoftwo}%
\providecommand \bibfield  [0]{\@secondoftwo}%
\providecommand \translation [1]{[#1]}%
\providecommand \BibitemOpen [0]{}%
\providecommand \bibitemStop [0]{}%
\providecommand \bibitemNoStop [0]{.\EOS\space}%
\providecommand \EOS [0]{\spacefactor3000\relax}%
\providecommand \BibitemShut  [1]{\csname bibitem#1\endcsname}%
\let\auto@bib@innerbib\@empty
\bibitem [{\citenamefont {Collaboration}\ \emph {et~al.}(2018)\citenamefont
  {Collaboration}, \citenamefont {Collaboration} \emph
  {et~al.}}]{ligo2018gwtc}%
  \BibitemOpen
  \bibfield  {author} {\bibinfo {author} {\bibfnamefont {L.~S.}\ \bibnamefont
  {Collaboration}}, \bibinfo {author} {\bibfnamefont {V.}~\bibnamefont
  {Collaboration}},  \emph {et~al.},\ }\href@noop {} {\bibfield  {journal}
  {\bibinfo  {journal} {arXiv preprint arXiv:1811.12907}\ } (\bibinfo {year}
  {2018})}\BibitemShut {NoStop}%
\bibitem [{\citenamefont {Aasi}\ \emph {et~al.}(2015)\citenamefont {Aasi},
  \citenamefont {Abbott}, \citenamefont {Abbott}, \citenamefont {Abbott},
  \citenamefont {Abernathy}, \citenamefont {Ackley}, \citenamefont {Adams},
  \citenamefont {Adams}, \citenamefont {Addesso}, \citenamefont {Adhikari}
  \emph {et~al.}}]{aasi2015advanced}%
  \BibitemOpen
  \bibfield  {author} {\bibinfo {author} {\bibfnamefont {J.}~\bibnamefont
  {Aasi}}, \bibinfo {author} {\bibfnamefont {B.}~\bibnamefont {Abbott}},
  \bibinfo {author} {\bibfnamefont {R.}~\bibnamefont {Abbott}}, \bibinfo
  {author} {\bibfnamefont {T.}~\bibnamefont {Abbott}}, \bibinfo {author}
  {\bibfnamefont {M.}~\bibnamefont {Abernathy}}, \bibinfo {author}
  {\bibfnamefont {K.}~\bibnamefont {Ackley}}, \bibinfo {author} {\bibfnamefont
  {C.}~\bibnamefont {Adams}}, \bibinfo {author} {\bibfnamefont
  {T.}~\bibnamefont {Adams}}, \bibinfo {author} {\bibfnamefont
  {P.}~\bibnamefont {Addesso}}, \bibinfo {author} {\bibfnamefont
  {R.}~\bibnamefont {Adhikari}},  \emph {et~al.},\ }\href@noop {} {\bibfield
  {journal} {\bibinfo  {journal} {Classical and quantum gravity}\ }\textbf
  {\bibinfo {volume} {32}},\ \bibinfo {pages} {074001} (\bibinfo {year}
  {2015})}\BibitemShut {NoStop}%
\bibitem [{\citenamefont {Evans}\ \emph {et~al.}(2010)\citenamefont {Evans},
  \citenamefont {Barsotti},\ and\ \citenamefont
  {Fritschel}}]{evans2010general}%
  \BibitemOpen
  \bibfield  {author} {\bibinfo {author} {\bibfnamefont {M.}~\bibnamefont
  {Evans}}, \bibinfo {author} {\bibfnamefont {L.}~\bibnamefont {Barsotti}}, \
  and\ \bibinfo {author} {\bibfnamefont {P.}~\bibnamefont {Fritschel}},\
  }\href@noop {} {\bibfield  {journal} {\bibinfo  {journal} {Physics Letters
  A}\ }\textbf {\bibinfo {volume} {374}},\ \bibinfo {pages} {665} (\bibinfo
  {year} {2010})}\BibitemShut {NoStop}%
\bibitem [{\citenamefont {Braginsky}\ \emph {et~al.}(2001)\citenamefont
  {Braginsky}, \citenamefont {Strigin},\ and\ \citenamefont
  {Vyatchanin}}]{braginsky2001parametric}%
  \BibitemOpen
  \bibfield  {author} {\bibinfo {author} {\bibfnamefont {V.}~\bibnamefont
  {Braginsky}}, \bibinfo {author} {\bibfnamefont {S.}~\bibnamefont {Strigin}},
  \ and\ \bibinfo {author} {\bibfnamefont {S.~P.}\ \bibnamefont {Vyatchanin}},\
  }\href@noop {} {\bibfield  {journal} {\bibinfo  {journal} {Physics Letters
  A}\ }\textbf {\bibinfo {volume} {287}},\ \bibinfo {pages} {331} (\bibinfo
  {year} {2001})}\BibitemShut {NoStop}%
\bibitem [{\citenamefont {Braginsky}\ \emph {et~al.}(2002)\citenamefont
  {Braginsky}, \citenamefont {Strigin},\ and\ \citenamefont
  {Vyatchanin}}]{braginsky2002analysis}%
  \BibitemOpen
  \bibfield  {author} {\bibinfo {author} {\bibfnamefont {V.~B.}\ \bibnamefont
  {Braginsky}}, \bibinfo {author} {\bibfnamefont {S.~E.}\ \bibnamefont
  {Strigin}}, \ and\ \bibinfo {author} {\bibfnamefont {S.~P.}\ \bibnamefont
  {Vyatchanin}},\ }\href@noop {} {\bibfield  {journal} {\bibinfo  {journal}
  {Physics Letters A}\ }\textbf {\bibinfo {volume} {305}},\ \bibinfo {pages}
  {111} (\bibinfo {year} {2002})}\BibitemShut {NoStop}%
\bibitem [{\citenamefont {Kells}\ and\ \citenamefont
  {D'Ambrosio}(2002)}]{kells2002considerations}%
  \BibitemOpen
  \bibfield  {author} {\bibinfo {author} {\bibfnamefont {W.}~\bibnamefont
  {Kells}}\ and\ \bibinfo {author} {\bibfnamefont {E.}~\bibnamefont
  {D'Ambrosio}},\ }\href@noop {} {\bibfield  {journal} {\bibinfo  {journal}
  {Physics Letters A}\ }\textbf {\bibinfo {volume} {299}},\ \bibinfo {pages}
  {326} (\bibinfo {year} {2002})}\BibitemShut {NoStop}%
\bibitem [{\citenamefont {Ju}\ \emph {et~al.}(2006)\citenamefont {Ju},
  \citenamefont {Gras}, \citenamefont {Zhao}, \citenamefont {Degallaix},\ and\
  \citenamefont {Blair}}]{ju2006multiple}%
  \BibitemOpen
  \bibfield  {author} {\bibinfo {author} {\bibfnamefont {L.}~\bibnamefont
  {Ju}}, \bibinfo {author} {\bibfnamefont {S.}~\bibnamefont {Gras}}, \bibinfo
  {author} {\bibfnamefont {C.}~\bibnamefont {Zhao}}, \bibinfo {author}
  {\bibfnamefont {J.}~\bibnamefont {Degallaix}}, \ and\ \bibinfo {author}
  {\bibfnamefont {D.}~\bibnamefont {Blair}},\ }\href@noop {} {\bibfield
  {journal} {\bibinfo  {journal} {Physics Letters A}\ }\textbf {\bibinfo
  {volume} {354}},\ \bibinfo {pages} {360} (\bibinfo {year}
  {2006})}\BibitemShut {NoStop}%
\bibitem [{\citenamefont {Gras}\ \emph {et~al.}(2010)\citenamefont {Gras},
  \citenamefont {Zhao}, \citenamefont {Blair},\ and\ \citenamefont
  {Ju}}]{gras2010parametric}%
  \BibitemOpen
  \bibfield  {author} {\bibinfo {author} {\bibfnamefont {S.}~\bibnamefont
  {Gras}}, \bibinfo {author} {\bibfnamefont {C.}~\bibnamefont {Zhao}}, \bibinfo
  {author} {\bibfnamefont {D.}~\bibnamefont {Blair}}, \ and\ \bibinfo {author}
  {\bibfnamefont {L.}~\bibnamefont {Ju}},\ }\href@noop {} {\bibfield  {journal}
  {\bibinfo  {journal} {Classical and Quantum Gravity}\ }\textbf {\bibinfo
  {volume} {27}},\ \bibinfo {pages} {205019} (\bibinfo {year}
  {2010})}\BibitemShut {NoStop}%
\bibitem [{\citenamefont {Vyatchanin}\ and\ \citenamefont
  {Strigin}(2012)}]{vyatchanin2012parametric}%
  \BibitemOpen
  \bibfield  {author} {\bibinfo {author} {\bibfnamefont {S.~P.}\ \bibnamefont
  {Vyatchanin}}\ and\ \bibinfo {author} {\bibfnamefont {S.~E.}\ \bibnamefont
  {Strigin}},\ }\href@noop {} {\bibfield  {journal} {\bibinfo  {journal}
  {Physics-Uspekhi}\ }\textbf {\bibinfo {volume} {55}},\ \bibinfo {pages}
  {1115} (\bibinfo {year} {2012})}\BibitemShut {NoStop}%
\bibitem [{\citenamefont {Evans}\ \emph {et~al.}(2015)\citenamefont {Evans},
  \citenamefont {Gras}, \citenamefont {Fritschel}, \citenamefont {Miller},
  \citenamefont {Barsotti}, \citenamefont {Martynov}, \citenamefont {Brooks},
  \citenamefont {Coyne}, \citenamefont {Abbott}, \citenamefont {Adhikari} \emph
  {et~al.}}]{evans2015observation}%
  \BibitemOpen
  \bibfield  {author} {\bibinfo {author} {\bibfnamefont {M.}~\bibnamefont
  {Evans}}, \bibinfo {author} {\bibfnamefont {S.}~\bibnamefont {Gras}},
  \bibinfo {author} {\bibfnamefont {P.}~\bibnamefont {Fritschel}}, \bibinfo
  {author} {\bibfnamefont {J.}~\bibnamefont {Miller}}, \bibinfo {author}
  {\bibfnamefont {L.}~\bibnamefont {Barsotti}}, \bibinfo {author}
  {\bibfnamefont {D.}~\bibnamefont {Martynov}}, \bibinfo {author}
  {\bibfnamefont {A.}~\bibnamefont {Brooks}}, \bibinfo {author} {\bibfnamefont
  {D.}~\bibnamefont {Coyne}}, \bibinfo {author} {\bibfnamefont
  {R.}~\bibnamefont {Abbott}}, \bibinfo {author} {\bibfnamefont {R.~X.}\
  \bibnamefont {Adhikari}},  \emph {et~al.},\ }\href@noop {} {\bibfield
  {journal} {\bibinfo  {journal} {Physical review letters}\ }\textbf {\bibinfo
  {volume} {114}},\ \bibinfo {pages} {161102} (\bibinfo {year}
  {2015})}\BibitemShut {NoStop}%
\bibitem [{\citenamefont {Zhao}\ \emph {et~al.}(2005)\citenamefont {Zhao},
  \citenamefont {Ju}, \citenamefont {Degallaix}, \citenamefont {Gras},\ and\
  \citenamefont {Blair}}]{zhao2005parametric}%
  \BibitemOpen
  \bibfield  {author} {\bibinfo {author} {\bibfnamefont {C.}~\bibnamefont
  {Zhao}}, \bibinfo {author} {\bibfnamefont {L.}~\bibnamefont {Ju}}, \bibinfo
  {author} {\bibfnamefont {J.}~\bibnamefont {Degallaix}}, \bibinfo {author}
  {\bibfnamefont {S.}~\bibnamefont {Gras}}, \ and\ \bibinfo {author}
  {\bibfnamefont {D.}~\bibnamefont {Blair}},\ }\href@noop {} {\bibfield
  {journal} {\bibinfo  {journal} {Physical review letters}\ }\textbf {\bibinfo
  {volume} {94}},\ \bibinfo {pages} {121102} (\bibinfo {year}
  {2005})}\BibitemShut {NoStop}%
\bibitem [{\citenamefont {Blair}\ \emph {et~al.}(2017)\citenamefont {Blair},
  \citenamefont {Gras}, \citenamefont {Abbott}, \citenamefont {Aston},
  \citenamefont {Betzwieser}, \citenamefont {Blair}, \citenamefont {DeRosa},
  \citenamefont {Evans}, \citenamefont {Frolov}, \citenamefont {Fritschel}
  \emph {et~al.}}]{blair2017first}%
  \BibitemOpen
  \bibfield  {author} {\bibinfo {author} {\bibfnamefont {C.}~\bibnamefont
  {Blair}}, \bibinfo {author} {\bibfnamefont {S.}~\bibnamefont {Gras}},
  \bibinfo {author} {\bibfnamefont {R.}~\bibnamefont {Abbott}}, \bibinfo
  {author} {\bibfnamefont {S.}~\bibnamefont {Aston}}, \bibinfo {author}
  {\bibfnamefont {J.}~\bibnamefont {Betzwieser}}, \bibinfo {author}
  {\bibfnamefont {D.}~\bibnamefont {Blair}}, \bibinfo {author} {\bibfnamefont
  {R.}~\bibnamefont {DeRosa}}, \bibinfo {author} {\bibfnamefont
  {M.}~\bibnamefont {Evans}}, \bibinfo {author} {\bibfnamefont
  {V.}~\bibnamefont {Frolov}}, \bibinfo {author} {\bibfnamefont
  {P.}~\bibnamefont {Fritschel}},  \emph {et~al.},\ }\href@noop {} {\bibfield
  {journal} {\bibinfo  {journal} {Physical review letters}\ }\textbf {\bibinfo
  {volume} {118}},\ \bibinfo {pages} {151102} (\bibinfo {year}
  {2017})}\BibitemShut {NoStop}%
\bibitem [{\citenamefont {Gras}\ \emph {et~al.}(2015)\citenamefont {Gras},
  \citenamefont {Fritschel}, \citenamefont {Barsotti},\ and\ \citenamefont
  {Evans}}]{gras2015resonant}%
  \BibitemOpen
  \bibfield  {author} {\bibinfo {author} {\bibfnamefont {S.}~\bibnamefont
  {Gras}}, \bibinfo {author} {\bibfnamefont {P.}~\bibnamefont {Fritschel}},
  \bibinfo {author} {\bibfnamefont {L.}~\bibnamefont {Barsotti}}, \ and\
  \bibinfo {author} {\bibfnamefont {M.}~\bibnamefont {Evans}},\ }\href@noop {}
  {\bibfield  {journal} {\bibinfo  {journal} {Physical Review D}\ }\textbf
  {\bibinfo {volume} {92}},\ \bibinfo {pages} {082001} (\bibinfo {year}
  {2015})}\BibitemShut {NoStop}%
\bibitem [{\citenamefont {Losurdo}\ \emph {et~al.}(1999)\citenamefont
  {Losurdo}, \citenamefont {Bernardini}, \citenamefont {Braccini},
  \citenamefont {Bradaschia}, \citenamefont {Casciano}, \citenamefont
  {Dattilo}, \citenamefont {De~Salvo}, \citenamefont {Di~Virgilio},
  \citenamefont {Frasconi}, \citenamefont {Gaddi} \emph
  {et~al.}}]{losurdo1999inverted}%
  \BibitemOpen
  \bibfield  {author} {\bibinfo {author} {\bibfnamefont {G.}~\bibnamefont
  {Losurdo}}, \bibinfo {author} {\bibfnamefont {M.}~\bibnamefont {Bernardini}},
  \bibinfo {author} {\bibfnamefont {S.}~\bibnamefont {Braccini}}, \bibinfo
  {author} {\bibfnamefont {C.}~\bibnamefont {Bradaschia}}, \bibinfo {author}
  {\bibfnamefont {C.}~\bibnamefont {Casciano}}, \bibinfo {author}
  {\bibfnamefont {V.}~\bibnamefont {Dattilo}}, \bibinfo {author} {\bibfnamefont
  {R.}~\bibnamefont {De~Salvo}}, \bibinfo {author} {\bibfnamefont
  {A.}~\bibnamefont {Di~Virgilio}}, \bibinfo {author} {\bibfnamefont
  {F.}~\bibnamefont {Frasconi}}, \bibinfo {author} {\bibfnamefont
  {A.}~\bibnamefont {Gaddi}},  \emph {et~al.},\ }\href@noop {} {\bibfield
  {journal} {\bibinfo  {journal} {Review of scientific instruments}\ }\textbf
  {\bibinfo {volume} {70}},\ \bibinfo {pages} {2507} (\bibinfo {year}
  {1999})}\BibitemShut {NoStop}%
\bibitem [{\citenamefont {Ballardin}\ \emph {et~al.}(2001)\citenamefont
  {Ballardin}, \citenamefont {Bracci}, \citenamefont {Braccini}, \citenamefont
  {Bradaschia}, \citenamefont {Casciano}, \citenamefont {Calamai},
  \citenamefont {Cavalieri}, \citenamefont {Cecchi}, \citenamefont {Cella},
  \citenamefont {Cuoco} \emph {et~al.}}]{ballardin2001measurement}%
  \BibitemOpen
  \bibfield  {author} {\bibinfo {author} {\bibfnamefont {G.}~\bibnamefont
  {Ballardin}}, \bibinfo {author} {\bibfnamefont {L.}~\bibnamefont {Bracci}},
  \bibinfo {author} {\bibfnamefont {S.}~\bibnamefont {Braccini}}, \bibinfo
  {author} {\bibfnamefont {C.}~\bibnamefont {Bradaschia}}, \bibinfo {author}
  {\bibfnamefont {C.}~\bibnamefont {Casciano}}, \bibinfo {author}
  {\bibfnamefont {G.}~\bibnamefont {Calamai}}, \bibinfo {author} {\bibfnamefont
  {R.}~\bibnamefont {Cavalieri}}, \bibinfo {author} {\bibfnamefont
  {R.}~\bibnamefont {Cecchi}}, \bibinfo {author} {\bibfnamefont
  {G.}~\bibnamefont {Cella}}, \bibinfo {author} {\bibfnamefont
  {E.}~\bibnamefont {Cuoco}},  \emph {et~al.},\ }\href@noop {} {\bibfield
  {journal} {\bibinfo  {journal} {Review of Scientific Instruments}\ }\textbf
  {\bibinfo {volume} {72}},\ \bibinfo {pages} {3643} (\bibinfo {year}
  {2001})}\BibitemShut {NoStop}%
\bibitem [{\citenamefont {Matichard}\ \emph {et~al.}(2015)\citenamefont
  {Matichard}, \citenamefont {Lantz}, \citenamefont {Mason}, \citenamefont
  {Mittleman}, \citenamefont {Abbott}, \citenamefont {Abbott}, \citenamefont
  {Allwine}, \citenamefont {Barnum}, \citenamefont {Birch}, \citenamefont
  {Biscans} \emph {et~al.}}]{matichard2015advanced}%
  \BibitemOpen
  \bibfield  {author} {\bibinfo {author} {\bibfnamefont {F.}~\bibnamefont
  {Matichard}}, \bibinfo {author} {\bibfnamefont {B.}~\bibnamefont {Lantz}},
  \bibinfo {author} {\bibfnamefont {K.}~\bibnamefont {Mason}}, \bibinfo
  {author} {\bibfnamefont {R.}~\bibnamefont {Mittleman}}, \bibinfo {author}
  {\bibfnamefont {B.}~\bibnamefont {Abbott}}, \bibinfo {author} {\bibfnamefont
  {S.}~\bibnamefont {Abbott}}, \bibinfo {author} {\bibfnamefont
  {E.}~\bibnamefont {Allwine}}, \bibinfo {author} {\bibfnamefont
  {S.}~\bibnamefont {Barnum}}, \bibinfo {author} {\bibfnamefont
  {J.}~\bibnamefont {Birch}}, \bibinfo {author} {\bibfnamefont
  {S.}~\bibnamefont {Biscans}},  \emph {et~al.},\ }\href@noop {} {\bibfield
  {journal} {\bibinfo  {journal} {Precision engineering}\ }\textbf {\bibinfo
  {volume} {40}},\ \bibinfo {pages} {287} (\bibinfo {year} {2015})}\BibitemShut
  {NoStop}%
\bibitem [{\citenamefont {Bergmann}\ \emph {et~al.}(2017)\citenamefont
  {Bergmann}, \citenamefont {Mow-Lowry}, \citenamefont {Adya}, \citenamefont
  {Bertolini}, \citenamefont {Hanke}, \citenamefont {Kirchhoff}, \citenamefont
  {K{\"o}hlenbeck}, \citenamefont {K{\"u}hn}, \citenamefont {Oppermann},
  \citenamefont {Wanner} \emph {et~al.}}]{bergmann2017passive}%
  \BibitemOpen
  \bibfield  {author} {\bibinfo {author} {\bibfnamefont {G.}~\bibnamefont
  {Bergmann}}, \bibinfo {author} {\bibfnamefont {C.}~\bibnamefont {Mow-Lowry}},
  \bibinfo {author} {\bibfnamefont {V.}~\bibnamefont {Adya}}, \bibinfo {author}
  {\bibfnamefont {A.}~\bibnamefont {Bertolini}}, \bibinfo {author}
  {\bibfnamefont {M.}~\bibnamefont {Hanke}}, \bibinfo {author} {\bibfnamefont
  {R.}~\bibnamefont {Kirchhoff}}, \bibinfo {author} {\bibfnamefont
  {S.}~\bibnamefont {K{\"o}hlenbeck}}, \bibinfo {author} {\bibfnamefont
  {G.}~\bibnamefont {K{\"u}hn}}, \bibinfo {author} {\bibfnamefont
  {P.}~\bibnamefont {Oppermann}}, \bibinfo {author} {\bibfnamefont
  {A.}~\bibnamefont {Wanner}},  \emph {et~al.},\ }\href@noop {} {\bibfield
  {journal} {\bibinfo  {journal} {Classical and Quantum Gravity}\ }\textbf
  {\bibinfo {volume} {34}},\ \bibinfo {pages} {065002} (\bibinfo {year}
  {2017})}\BibitemShut {NoStop}%
\bibitem [{\citenamefont {Sodano}\ \emph {et~al.}(2004)\citenamefont {Sodano},
  \citenamefont {Park},\ and\ \citenamefont {Inman}}]{sodano2004estimation}%
  \BibitemOpen
  \bibfield  {author} {\bibinfo {author} {\bibfnamefont {H.~A.}\ \bibnamefont
  {Sodano}}, \bibinfo {author} {\bibfnamefont {G.}~\bibnamefont {Park}}, \ and\
  \bibinfo {author} {\bibfnamefont {D.}~\bibnamefont {Inman}},\ }\href@noop {}
  {\bibfield  {journal} {\bibinfo  {journal} {Strain}\ }\textbf {\bibinfo
  {volume} {40}},\ \bibinfo {pages} {49} (\bibinfo {year} {2004})}\BibitemShut
  {NoStop}%
\bibitem [{\citenamefont {Hagood}\ and\ \citenamefont {von
  Flotow}(1991)}]{hagood1991damping}%
  \BibitemOpen
  \bibfield  {author} {\bibinfo {author} {\bibfnamefont {N.~W.}\ \bibnamefont
  {Hagood}}\ and\ \bibinfo {author} {\bibfnamefont {A.}~\bibnamefont {von
  Flotow}},\ }\href@noop {} {\bibfield  {journal} {\bibinfo  {journal} {Journal
  of Sound and Vibration}\ }\textbf {\bibinfo {volume} {146}},\ \bibinfo
  {pages} {243} (\bibinfo {year} {1991})}\BibitemShut {NoStop}%
\bibitem [{\citenamefont {Jaffe}(2012)}]{jaffe2012piezoelectric}%
  \BibitemOpen
  \bibfield  {author} {\bibinfo {author} {\bibfnamefont {B.}~\bibnamefont
  {Jaffe}},\ }\href@noop {} {\emph {\bibinfo {title} {Piezoelectric
  ceramics}}},\ Vol.~\bibinfo {volume} {3}\ (\bibinfo  {publisher} {Elsevier},\
  \bibinfo {year} {2012})\BibitemShut {NoStop}%
\bibitem [{\citenamefont {Bansal}\ and\ \citenamefont
  {Doremus}(2013)}]{bansal2013handbook}%
  \BibitemOpen
  \bibfield  {author} {\bibinfo {author} {\bibfnamefont {N.~P.}\ \bibnamefont
  {Bansal}}\ and\ \bibinfo {author} {\bibfnamefont {R.~H.}\ \bibnamefont
  {Doremus}},\ }\href@noop {} {\emph {\bibinfo {title} {Handbook of glass
  properties}}}\ (\bibinfo  {publisher} {Elsevier},\ \bibinfo {year}
  {2013})\BibitemShut {NoStop}%
\bibitem [{\citenamefont {Zener}(1940)}]{zener1940internal}%
  \BibitemOpen
  \bibfield  {author} {\bibinfo {author} {\bibfnamefont {C.}~\bibnamefont
  {Zener}},\ }\href@noop {} {\bibfield  {journal} {\bibinfo  {journal}
  {Proceedings of the Physical Society}\ }\textbf {\bibinfo {volume} {52}},\
  \bibinfo {pages} {152} (\bibinfo {year} {1940})}\BibitemShut {NoStop}%
\bibitem [{\citenamefont {Biscans}\ \emph {et~al.}(2018)\citenamefont
  {Biscans}, \citenamefont {Gras}, \citenamefont {Evans}, \citenamefont
  {Fritschel}, \citenamefont {Pezerat},\ and\ \citenamefont
  {Picart}}]{biscans2018method}%
  \BibitemOpen
  \bibfield  {author} {\bibinfo {author} {\bibfnamefont {S.}~\bibnamefont
  {Biscans}}, \bibinfo {author} {\bibfnamefont {S.}~\bibnamefont {Gras}},
  \bibinfo {author} {\bibfnamefont {M.}~\bibnamefont {Evans}}, \bibinfo
  {author} {\bibfnamefont {P.}~\bibnamefont {Fritschel}}, \bibinfo {author}
  {\bibfnamefont {C.}~\bibnamefont {Pezerat}}, \ and\ \bibinfo {author}
  {\bibfnamefont {P.}~\bibnamefont {Picart}},\ }\href@noop {} {\bibfield
  {journal} {\bibinfo  {journal} {Journal of Sound and Vibration}\ }\textbf
  {\bibinfo {volume} {423}},\ \bibinfo {pages} {118} (\bibinfo {year}
  {2018})}\BibitemShut {NoStop}%
\bibitem [{\citenamefont {Biscans}(2018)}]{biscans2018optimization}%
  \BibitemOpen
  \bibfield  {author} {\bibinfo {author} {\bibfnamefont {S.}~\bibnamefont
  {Biscans}},\ }\emph {\bibinfo {title} {Optimization of the Advanced LIGO
  gravitational-wave detectors duty cycle by reduction of parametric
  instabilities and environmental impacts}},\ \href@noop {} {Ph.D. thesis},\
  \bibinfo  {school} {Le Mans} (\bibinfo {year} {2018})\BibitemShut {NoStop}%
\bibitem [{ans()}]{ansys}%
  \BibitemOpen
  \href@noop {} {\emph {\bibinfo {title} {Ansys academic research, release
  17.1}}}\BibitemShut {NoStop}%
\bibitem [{\citenamefont {Callen}\ and\ \citenamefont
  {Welton}(1951)}]{callen1951irreversibility}%
  \BibitemOpen
  \bibfield  {author} {\bibinfo {author} {\bibfnamefont {H.~B.}\ \bibnamefont
  {Callen}}\ and\ \bibinfo {author} {\bibfnamefont {T.~A.}\ \bibnamefont
  {Welton}},\ }\href@noop {} {\bibfield  {journal} {\bibinfo  {journal}
  {Physical Review}\ }\textbf {\bibinfo {volume} {83}},\ \bibinfo {pages} {34}
  (\bibinfo {year} {1951})}\BibitemShut {NoStop}%
\bibitem [{\citenamefont {Levin}(1998)}]{levin1998internal}%
  \BibitemOpen
  \bibfield  {author} {\bibinfo {author} {\bibfnamefont {Y.}~\bibnamefont
  {Levin}},\ }\href@noop {} {\bibfield  {journal} {\bibinfo  {journal}
  {Physical Review D}\ }\textbf {\bibinfo {volume} {57}},\ \bibinfo {pages}
  {659} (\bibinfo {year} {1998})}\BibitemShut {NoStop}%
\bibitem [{\citenamefont {Martynov}\ \emph {et~al.}(2017)\citenamefont
  {Martynov}, \citenamefont {Frolov}, \citenamefont {Kandhasamy}, \citenamefont
  {Izumi}, \citenamefont {Miao}, \citenamefont {Mavalvala}, \citenamefont
  {Hall}, \citenamefont {Lanza}, \citenamefont {Abbott}, \citenamefont {Abbott}
  \emph {et~al.}}]{martynov2017quantum}%
  \BibitemOpen
  \bibfield  {author} {\bibinfo {author} {\bibfnamefont {D.~V.}\ \bibnamefont
  {Martynov}}, \bibinfo {author} {\bibfnamefont {V.}~\bibnamefont {Frolov}},
  \bibinfo {author} {\bibfnamefont {S.}~\bibnamefont {Kandhasamy}}, \bibinfo
  {author} {\bibfnamefont {K.}~\bibnamefont {Izumi}}, \bibinfo {author}
  {\bibfnamefont {H.}~\bibnamefont {Miao}}, \bibinfo {author} {\bibfnamefont
  {N.}~\bibnamefont {Mavalvala}}, \bibinfo {author} {\bibfnamefont
  {E.}~\bibnamefont {Hall}}, \bibinfo {author} {\bibfnamefont {R.}~\bibnamefont
  {Lanza}}, \bibinfo {author} {\bibfnamefont {B.}~\bibnamefont {Abbott}},
  \bibinfo {author} {\bibfnamefont {R.}~\bibnamefont {Abbott}},  \emph
  {et~al.},\ }\href@noop {} {\bibfield  {journal} {\bibinfo  {journal}
  {Physical Review A}\ }\textbf {\bibinfo {volume} {95}},\ \bibinfo {pages}
  {043831} (\bibinfo {year} {2017})}\BibitemShut {NoStop}%
\bibitem [{\citenamefont {Cahillane}\ \emph {et~al.}(2017)\citenamefont
  {Cahillane}, \citenamefont {Betzwieser}, \citenamefont {Brown}, \citenamefont
  {Goetz}, \citenamefont {Hall}, \citenamefont {Izumi}, \citenamefont
  {Kandhasamy}, \citenamefont {Karki}, \citenamefont {Kissel}, \citenamefont
  {Mendell}, \citenamefont {Savage}, \citenamefont {Tuyenbayev}, \citenamefont
  {Urban}, \citenamefont {Viets}, \citenamefont {Wade},\ and\ \citenamefont
  {Weinstein}}]{ligoc2017alibration}%
  \BibitemOpen
  \bibfield  {author} {\bibinfo {author} {\bibfnamefont {C.}~\bibnamefont
  {Cahillane}}, \bibinfo {author} {\bibfnamefont {J.}~\bibnamefont
  {Betzwieser}}, \bibinfo {author} {\bibfnamefont {D.~A.}\ \bibnamefont
  {Brown}}, \bibinfo {author} {\bibfnamefont {E.}~\bibnamefont {Goetz}},
  \bibinfo {author} {\bibfnamefont {E.~D.}\ \bibnamefont {Hall}}, \bibinfo
  {author} {\bibfnamefont {K.}~\bibnamefont {Izumi}}, \bibinfo {author}
  {\bibfnamefont {S.}~\bibnamefont {Kandhasamy}}, \bibinfo {author}
  {\bibfnamefont {S.}~\bibnamefont {Karki}}, \bibinfo {author} {\bibfnamefont
  {J.~S.}\ \bibnamefont {Kissel}}, \bibinfo {author} {\bibfnamefont
  {G.}~\bibnamefont {Mendell}}, \bibinfo {author} {\bibfnamefont {R.~L.}\
  \bibnamefont {Savage}}, \bibinfo {author} {\bibfnamefont {D.}~\bibnamefont
  {Tuyenbayev}}, \bibinfo {author} {\bibfnamefont {A.}~\bibnamefont {Urban}},
  \bibinfo {author} {\bibfnamefont {A.}~\bibnamefont {Viets}}, \bibinfo
  {author} {\bibfnamefont {M.}~\bibnamefont {Wade}}, \ and\ \bibinfo {author}
  {\bibfnamefont {A.~J.}\ \bibnamefont {Weinstein}},\ }\href {\doibase
  10.1103/PhysRevD.96.102001} {\bibfield  {journal} {\bibinfo  {journal} {Phys.
  Rev. D}\ }\textbf {\bibinfo {volume} {96}},\ \bibinfo {pages} {102001}
  (\bibinfo {year} {2017})}\BibitemShut {NoStop}%
\end{thebibliography}%

\end{document}